\documentclass{emulateapj}
\usepackage{hyperref}
\hypersetup{colorlinks=true, citecolor=blue}
\usepackage{graphicx}
\usepackage{txfonts}
\usepackage{color}
\newcommand{\peg}{{\sc P\'egase II} }
\def\Tiny{\fontsize{5pt}{5pt}\selectfont}

\begin{document}
\title{Star Formation in Collision Debris: Insights from the modeling of their Spectral Energy Distribution}

\author{M. Boquien\altaffilmark{1}, P.-A. Duc\altaffilmark{2}, F. Galliano\altaffilmark{2}, J. Braine\altaffilmark{3}, U. Lisenfeld\altaffilmark{4}, V. Charmandaris\altaffilmark{5,6,7}, P. N. Appleton\altaffilmark{8}}
\email{boquien@astro.umass.edu}

\altaffiltext{1}{University of Massachusetts, Department of Astronomy, LGRT-B 619E, Amherst, MA 01003, USA}
\altaffiltext{2}{Laboratoire AIM, CEA/DSM-CNRS-Universit\'e Paris Diderot, IRFU/Service d'Astrophysique, CEA-Saclay, 91191 Gif-sur-Yvette Cedex, France}
\altaffiltext{3}{Laboratoire d'Astrophysique de Bordeaux, OASU, Universit\'e Bordeaux 1, UMR 5804, CNRS/INSU, B.P. 89, F-33270 Floirac, France}
\altaffiltext{4}{Dept. de F\'\i sica Te\'orica y del Cosmos, Universidad de Granada, Granada, Spain}
\altaffiltext{5}{Department of Physics, University of Crete, GR-71003, Heraklion, Greece}
\altaffiltext{6}{IESL/Foundation for Research \& Technology-Hellas, GR-71110, Heraklion, Greece}
\altaffiltext{7}{Chercheur Associ\'e, Observatoire de Paris, F-75014, Paris, France}
\altaffiltext{8}{NASA Herschel Science Center, California Institute of Technology, Mail code 100-22, 770 South Wilson Ave., Pasadena, CA 91125, USA}

\begin{abstract}
During galaxy-galaxy interactions, massive gas clouds can be injected into the intergalactic medium which in turn become gravitationally bound, collapse and form stars, star clusters or even dwarf galaxies. The objects resulting from this process are both ``pristine'', as they are forming their first generation of stars, and chemically evolved because the metallicity inherited from their parent galaxies is high. Such characteristics make them particularly interesting laboratories to study star formation.
After having investigated their star-forming properties, we use photospheric, nebular and dust modeling to analyze here their spectral energy distribution (SED) from the far-ultraviolet to the mid-infrared regime for a sample of 7 star-forming regions.
Our analysis confirms that the intergalactic star forming regions in Stephan's Quintet, around Arp~105, and NGC~5291, appear devoid of stellar populations older than $10^9$~years. We also find an excess of light in the near-infrared regime (from 2~$\mu$m to 4.5~$\mu$m) which cannot be attributed to stellar photospheric or nebular contributions. This excess is correlated with the star formation rate intensity suggesting that it is probably due to emission by very small grains fluctuating in temperature as well as the polycyclic aromatic hydrocarbons (PAH) line at 3.3~$\mu$m. Comparing the attenuation via the Balmer decrement to the mid-infrared emission allows us to check the reliability of the attenuation estimate. It suggests the presence of embedded star forming regions in NGC~5291 and NGC~7252. Overall the SED of star-forming regions in collision debris (and Tidal Dwarf Galaxies) resemble more that of dusty star-forming regions in galactic disks than to that of typical star-forming dwarf galaxies. 
\end{abstract}


\section{Introduction}
\label{sec:intro}

Collision debris are the offsprings of close encounters between two or more galaxies. Their characteristics such as their morphologies or contents can vary greatly depending on the parameters of the interaction (prograde/retrograde, relative speed, mass ratio, galaxy type, impact parameter, etc.): they might appear as long, narrow tails, diffuse bridges, shells and rings \citep{schombert1990a}. The smallest objects in collision debris are the size of HII regions, many of which have been discovered recently between galaxies in nearby groups and clusters \citep{gerhard2002a,yoshida2002a,sakai2002a,weilbacher2003a,ryan2004a,mendes2004a,cortese2006a,werk2008a,werk2010a}. These HII regions, also known as Emission Line Dots for their compact appearance, have a relatively high metallicity, consistent with the idea that the gas in which they lie was somehow ejected from parent galaxies. Larger objects are often found in the tails and filaments resulting from tidal interactions between (gas-rich) galaxies. In the $10^6$~M$_\sun$ to $10^7$~M$_\sun$ mass range, some super star clusters can be formed \citep{weilbacher2003a,degrijs2003a,lopez2004a}. The largest HII regions are typically found at the apparent tip of tidal tails, at more than 100~kpc from the parent galaxies, in maxima of the HI column density where we also detect large quantities of molecular gas \citep{braine2000a,braine2001a,lisenfeld2002a} and where enough collision debris are often available to build a Tidal Dwarf Galaxy \citep[TDG,  ``an object which is a self-gravitating entity, formed out of the debris of a gravitational interaction'',][]{duc2000a}. All these regions have the exceptional characteristic of being intergalactic star-forming regions and they are embedded in HI clouds out of which the stars have formed recently.

Across their full range of masses, those debris -- especially the ones that are gas-rich and host star-forming regions -- have generated an increasing interest in the astronomical community over the past two decades. Indeed, for some of them the lack of a substantial evolved stellar population originating from the parent galaxies -- that is, they are practically a single stellar population -- makes them attractive targets to study star formation outside galaxies \citep[][for an assessment of their use as laboratories to study star formation]{boquien2009a} to address a variety of problems such as constraining the parameters of star formation and they also give hope to constrain the Initial Mass Function. At the same time, their high metallicity, inherited from their parent galaxies, makes the study of their interstellar medium -- in particular the detection of molecular gas through CO emission and dust -- easier than in metal deficient dwarf galaxies.

Studies of collision debris have addressed their molecular gas content \citep{braine2000a,braine2001a,lisenfeld2002a,lisenfeld2004a},
their dust emission \citep{higdon2006a,smith2007a,boquien2009a}, star formation rates \citep{hancock2007a,boquien2007a,smith2008a,boquien2009a}, stellar populations \citep{weilbacher2000a,mendes2004a,demello2008a,smith2008a,werk2008a,hancock2009a,sheen2009a}, dynamical properties \citep{bournaud2003a,bournaud2004a,duc2004a,mendes2006a,amram2007a} and also dark matter content \citep{bournaud2007a,milgrom2007a}.

The recent availability of images of interacting galaxies by GALEX, Spitzer and Akari allows us to construct the SED (spectral energy distribution) of collision debris, with data points ranging from the far-UV (ultraviolet) to the mid-IR (infrared). Fitting these SED with theoretical models, one may in principle determine several key parameters, such as the star formation rate, the star formation history, the attenuation as well as the total stellar and dust mass (taking into account the PAH [polycyclic aromatic hydrocarbons], the silicates and the graphites). In this paper, we present the first consistent UV to IR modeling of a sample of star-forming collision debris. Seven regions selected from the sample of intergalactic star-forming regions by \cite{boquien2009a} are studied. 

A number of questions remain open regarding star forming regions in the intergalactic medium, that can be addressed through such SED modeling. Are the star formation rate estimators usually used for spiral galaxies \citep{kennicutt1998a} reliable in intergalactic star-forming regions? Are there really collision debris that are constituted of a single, young,  stellar population, that is that lack a billion years old stellar component? If so, can such simple regions be used to constrain the IMF (initial mass function) in an easier way than in other environments?  What is the origin of their strong near-IR emission (as measured in the Spitzer 3.6~$\mu$m and 4.5~$\mu$m bands) if not due to an evolved stellar population?

The sample and data are presented in section 2; the modeling of their spectral energy distribution is described in section 3; we present the results in section 4; we discuss them in section 5 and finally we conclude in section 6.

\section{Sample and data}
\label{sec:sample}

\subsection{Sample}

Here our aim is to gain insights on star formation in intergalactic star-forming regions forming in collision debris, by modeling their spectral energy distribution. To do so, we have assembled a sample of intergalactic star-forming regions in interacting systems. The sample is drawn from the catalog of star forming regions in collision debris published by \cite{boquien2009a}. The objects in this catalog are specifically selected to span over a large range of star formation rates (from 0.002~M$_\sun$~yr$^{-1}$ to 0.2~M$_\sun$~yr$^{-1}$ as measured in the ultraviolet) in order to probe any variation from that of intergalactic HII regions to that of candidate TDG. All the regions selected in the present paper have been confirmed to be in the intergalactic medium thanks to observations (HI, CO, UV, optical, IR, spectroscopy) but also thanks to numerical simulations for a number of systems. Moreover, spectroscopic observations show they they all have a relatively high metallicity ($8.4\leq12+\log O/H\leq8.8$), confirming that they are made from pre-enriched gas, and thus most likely from collision debris. In addition, we have also restricted the sample to those having high enough signal to noise data points in the UV, optical and mid-infrared wavelength domains, necessary for the modeling of their SED. Finally we have also made sure that all regions are dominated by photoionization processes and not shocks as we focus on star formation. This was performed measuring line ratios in available spectra of HII regions and comparing with photoionization models. This was particularly crucial for Stephan's Quintet which contains some regions that are strongly affected by shocks. These selection criteria give us a sample of 7 objects that will be studied in this paper.

\begin{figure*}[htbp]
\includegraphics[height=\textwidth,angle=-90]{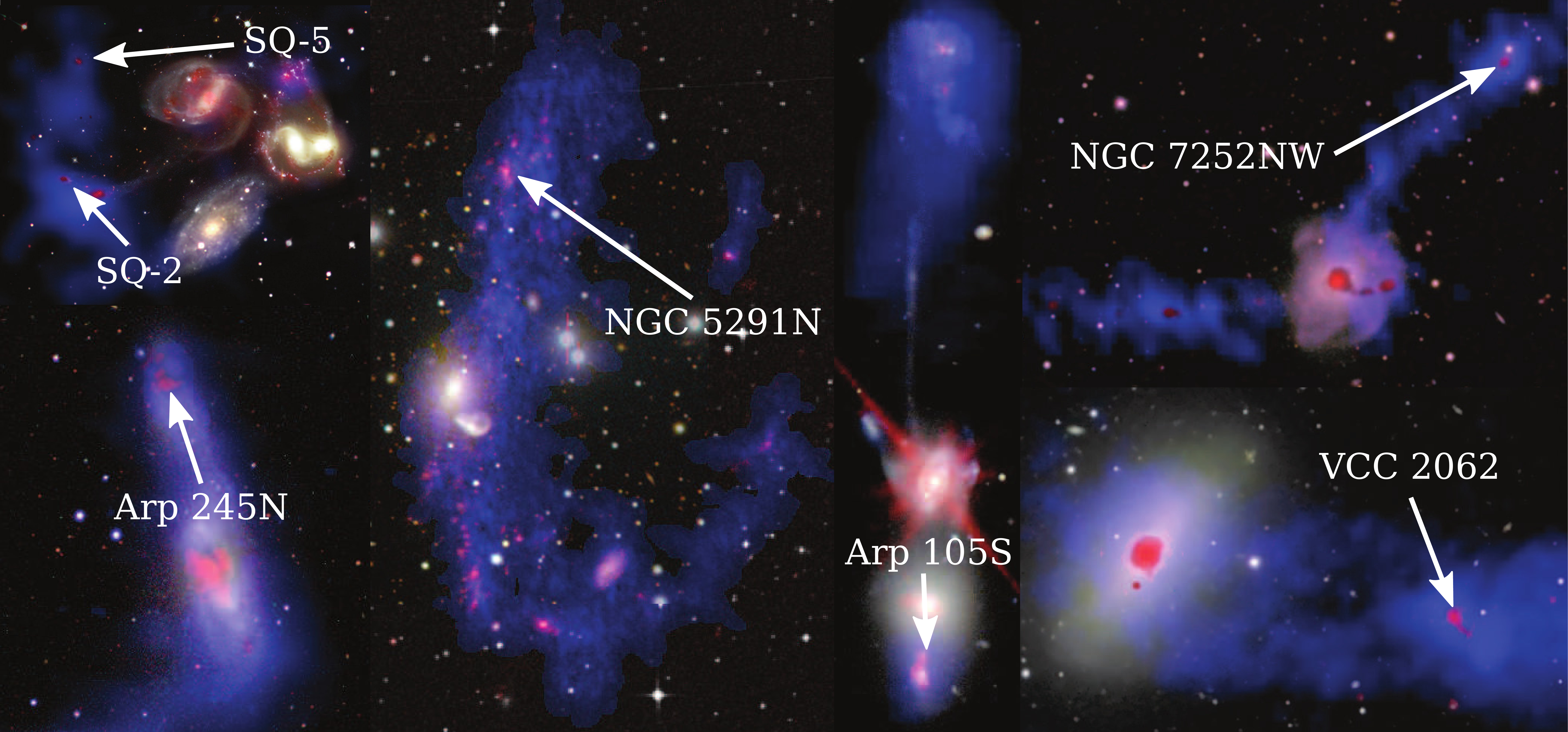}
\caption{True-color images of the selected systems. In addition, the neutral gas is overlayed in blue and star formation through FUV, H$\alpha$ or the 8~$\mu$m, is overlayed in red. More specifically in Stephan's Quintet: VLA/HI (blue), Spitzer/IRAC 8~$\mu$m (red), HST+Calar Alto 3.5~m (BVR). Arp~245: VLA-C+B/HI (blue), GALEX/FUV (red), Dupont 2.54m (UVI). NGC~5291: VLA-B/HI (blue), GALEX/FUV (red), DSS+NTT (BVR). Arp~105: VLA-C+B/HI (blue), Spitzer/IRAC 8~$\mu$m (red), CFHT/12k (BVR). NGC~7252: VLA-D/HI (blue), CFHT/H$\alpha$ (red), DSS+NTT (BVR). VCC~2062: VLA-C/HI (blue), GALEX/FUV (red), optical SDSS+NTT (BVR). North is up, East is left.}
\label{fig:systems}
\end{figure*}

All systems are presented in Figure~\ref{fig:systems}. The first two regions we have selected belong to Stephan's Quintet. The first one, named SQ-2 in \cite{boquien2009a} is located at the tip of a gaseous+stellar tidal tail in the south-eastern part of the system. The second one, named SQ-5 is located in the HI tail in the north-east of the system. It is a low mass compact region with blue optical colors, resembling a series of point sources with strong spectral line emission. Arp~105 is a bona fide interacting system between a spiral galaxy and an elliptical one. There are two prominent tidal tails. The one towards the north is a mix of stars and gas. The one towards the south is likely constituted only of gas which makes it particularly interesting, with Arp~105S a star forming region at its tip. Arp~245 is an interacting system with a large stellar and gaseous tidal tail towards the north. We have selected Arp~245N which is a star forming region at its extremity. NGC~5291 is a system constituted of a large collisional ring containing numerous star-forming regions. NGC~5291N is the brightest star-forming region in the ring and the brightest intergalactic star-forming region known so far in the ultraviolet, which is why we have selected it. Well-known system and typical of the last stage of the \cite{toomre1977a} merging sequence, NGC~7252 has two very extended tidal tails. We have selected NGC~7252NW, a massive star-forming region at the extremity of one of the tails. Finally we have selected VCC~2062 which is a probably the most evolved intergalactic region known to date but still forming stars \citep{duc2007a}. These regions enable us to probe a large range of SFR in the intergalactic medium allowing us to investigate changes of the dust SED and the contribution of the evolved stellar populations to the infrared emission as a function of SFR.

\subsection{Data: from the UV to the mid-IR}

\begin{itemize}
\item All ultraviolet GALEX and H$\alpha$ CFHT/ESO data fluxes are published in \cite{boquien2009a}. 

\item Optical broad--band images of Arp~105, Arp~245, NGC~5291 and VCC~2062 are respectively published in \cite{duc1994a}, \cite{duc2000a}, \cite{duc1998a} and \cite{duc2007a}. Optical B, V and R-band images of Stephan's Quintet and NGC~7252 were obtained by our group using the 3.5m at Calar Alto observatory and the NTT at La Silla ESO observatory.

\item Near-infrared images of NGC~7252 and Arp~245 were obtained with the Akari space telescope and are published here for the first time. The N2 filter centered at 2.4~$\mu$m was used. Further ground-based near-infrared, J, H and K images of Arp~105, Arp~245 and NGC~5291 are published by \cite{duc1994a,duc1998a,duc2000a}. 

\item Mid--infrared Spitzer/IRAC data fluxes in the 3.6, 4.5 and 8 $\mu$m bands are published in \cite{boquien2009a}. Spitzer observations performed with the Short-Low module of IRS between 5 and 15  $\mu$m  are available for Arp~105S, Arp~245N \citep{boquien2009a} and NGC~5291N \citep{higdon2006a}. A 24~$\mu$m image of Stephan's Quintet obtained with Spitzer/MIPS is presented in \cite{cluver2010a}. 24~$\mu$m images of NGC~5291 and NGC~7252 were obtained as part of our Spitzer's GO program 50191 and are published here for the first time. The VCC~2062 24~$\mu$m image was obtained from the public Spitzer archives.

\end{itemize}
For all these images, aperture photometry has been carried out following the procedures described in \cite{boquien2009a}. The apertures correspond to the ones presented in the aforementioned article, slightly widened if necessary to enclose the 24~$\mu$m flux. For convenience we list in Tab.~\ref{tab:fluxes} (Sec.~\ref{sec:results}) the fluxes that we use in the present paper.

Optical spectroscopy is available for all selected star forming regions but one. The slits are more narrow than the apertures and focus on the brightest parts of each region. The metallicity and attenuation of the ionized gas are published in the aforementioned articles. We have summarized the observed parameters in Table~\ref{tab:spec-parameters}, Sec.~\ref{sec:results}.

\section{Spectral energy distribution modeling}

In this section we model the spectral energy distribution of 7 intergalactic star forming regions using a photospheric+nebular model and a dust model.

We divide the spectral energy distribution modeling into two steps. In the first step we use {\sc P\'egase II} to fit the continuum and line emission from the stars and the gas using observations ranging from the ultraviolet to the near-infrared. The output of this fit provides information on the excitation and intrinsic interstellar radiation field. In the second step we use those estimates to fit the mid-infrared observations using a dust model. In the process we also estimate the energy absorbed and re-emitted by the dust at longer wavelengths.

\subsection{Ultraviolet to near-infrared spectral energy distribution modeling}

We combine spectroscopic observations of emission lines with the \peg spectral evolution code \citep{fioc1997a} to model star forming regions in collision debris. 

We use the \cite{salpeter1955a} IMF from 0.1~M$_\sun$ to 100~M$_\sun$ as a baseline for this study unless specified otherwise. 

\subsubsection{Star formation history}

It is known that gaseous disks in parent galaxies are often more extended than the stellar ones. As a result, depending on the parameters of the interaction, the collision debris may be devoid of any older pre-existing evolved stellar population, or on the contrary be dominated by old stars. We model this using a two burst star formation history. The first burst models the disk of a spiral galaxy, that is the evolved stellar population that is stripped during an interaction: it assumes a $12\times10^9$ years old exponentially declining burst which has consumed 90\% of its original gas reservoir. On top of this evolved population we add a second burst which models the star formation episode taking place in the collision debris after $12\times10^9$ years. As the photometric properties change very rapidly during the first few million years the burst is modeled by an exponentially decreasing SFR: $SFR(t)=a\times\exp\left(-t/\tau\right)$, where $a$ determines the intensity of the burst and $\tau$ is the timescale. We follow this second burst from $t=10^6$ years to $t=2\times10^9$ years, $t$ being here the time elapsed since the beginning of the star formation episode in collision debris. For this population, $\tau$ varies between $10^6$ years and $10\times10^9$ years to span a large range of possible star formation histories from nearly instantaneous to nearly constant.

The quality of the determination of the relative contribution of both components will ascertain the accuracy of the determination of the other parameters. The relative importance of the two populations is determined by the ratio $r$, the stellar mass produced by the second burst divided by the mass of the stars produced by the first burst (the evolved population), at a given time $t$ for a given timescale $\tau$. We span a range of ratios from $\log r=-2.75$ to $\log r=2$ by steps of $0.25$. The parameter $r$ can be seen as a burst strength.

A possible limitation is the imprecise modeling of TP-AGB (thermally pulsating asymptotic giant branch) stars by \peg. \cite{maraston2005a} has shown that they are dominant for populations aged between $300\times10^6$ and $2\times10^9$ years. For our specific case however, the older population is much older and is dominated by red giant branch stars and the younger populations best fit is never older than $300\times10^6$ years old as we will show later.

\subsubsection{Nebular emission}

The nebular emission accounts for a significant fraction of the flux in the broad optical bands from young star forming regions \citep{anders2003a}. To improve the standard nebular modeling of emission lines by \peg -- which assumes a given set of ratios between emission lines -- we use attenuation-corrected emission lines obtained from spectra. To calculate the modeled flux in each line, we scale the observed spectrum so that the observed H$\beta$ flux matches the modeled one. This is calculated using the following relation: $F\left(H\beta\right)=4.757\times10^{-20}\times N_{Lyc}\times f$, where $F\left(H\beta\right)$ is the flux in W, $N_{Lyc}$ is the number of Lyman continuum photons and $f$ the fraction of Lyman continuum photons ionizing an hydrogen atom usually taken as $0.7$ \citep{mezger1978a}. When there is no spectrum available, most notably in the near-infrared we keep the default \peg lines ratios. The nebular continuum is also modeled using \peg prescriptions. The attenuation of the lines is performed separately in a subsequent step (Sect.~\ref{sssec:extinction}).

\subsubsection{Metallicity}

The metallicity has a significant influence on the stellar spectrum. Indeed, a low metallicity star is hotter and emits more ionizing photons, affecting the strength of the emission lines and star formation tracing bands. The star forming regions selected in the sample have a relatively high metallicity as shown by spectroscopic observations from the literature \citep[][see Table~\ref{tab:spec-parameters}]{duc1994a,duc1998a,duc2000a,duc2007a}. We have assumed that the metallicity of the stars and the gas is constant. While local enrichment may be particularly important at low metallicity, the effect is weaker in collision debris because of their high metallicity. We have therefore set the metallicity to the value observed spectroscopically.

\subsubsection{Extinction\label{sssec:extinction}}

Even though \peg has provisions to handle the attenuation, we have performed this separately. We have modeled the attenuation of the burst taking place in the collision debris using the starburst attenuation law \citep{calzetti2001a}, a standard law for strongly star-forming regions. Another law such as the SMC (Small Magellanic Cloud) law has a much stronger slope in the ultraviolet for a given $A_V$. However, such an attenuation law is unlikely as star forming regions in collision debris have a metallicity significantly higher than the SMC one (see Table~\ref{tab:spec-parameters}). For the evolved population which has been stripped from the parent galaxies, we rather use the LMC (Large Magellanic Cloud) attenuation law \citep{gordon2003a} with a foreground screen geometry. This law exhibits the typical 220~nm bump found in a number of spiral galaxies which is suitable for the evolved stellar population originating from the parent galaxies. In addition, this population shows up mostly in the optical and in the near-infrared part of the SED where differences between dust attenuation laws are not crucial. Thus, the choice of a different law for this population would only have a very small impact. For each population, we span independently an attenuation range from $A_V=0$ to $A_V=2.9$ by steps of 0.1 magnitude.

The separation between the attenuation of both components allows a simpler interpretation of the results and avoids introducing new parameters to characterize the distribution of the dust and of the different stellar populations. In addition as star--forming regions are generally more extinguished than regions dominated by an evolved stellar population and are also much more compact, the error should remain minimal.

\subsubsection{Minimization}

To fit the models to the observations, we determine which one has the smallest $\chi^2$:

\begin{equation}
 \chi^2\left(a_1,\cdots,a_M\right)=\sum_{i=1}^N\left[\frac{y_i-\alpha\times m_i\left(a_1,\cdots,a_M\right)}{\sigma_i}\right]^2,
\end{equation}
where $a_j$ are the values of the parameters of the model, $M$ is the number of parameters of the model, $y_i$ are the observed fluxes with uncertainties $\sigma_i$ (assumed random, independent and distributed according to a Gaussian law), $N$ is the number of measurement points, $m_i$ are the fluxes of the model to test, and $\alpha$ the following normalization factor:

\begin{equation}
 \alpha=\frac{\sum_{i=1}^Ny_i\times m_i\left(a_1,\cdots,a_M\right)/\sigma_i^2}{\sum_{i=1}^Nm_i\left(a_1,\cdots,a_M\right)^2/\sigma_i^2}.
\end{equation}
All models produced by \peg are normalized to 1~M$_\sun$ of gas. This normalization factor allows us to directly scale the model to the observations.

This minimization is only performed on stellar population sensitive bands from FUV to near-infrared.

\subsubsection{Summary of free parameters}

The models have 5 free parameters: 1) the age $t$ of the burst of star formation in collision debris, ranging from $t=10^6$ years and $t=2\times10^9$ years, 2) the time constant $\tau$ of this burst ranging from $\tau=10^6$ years (nearly instantaneous) to $\tau=10\times10^9$ years (nearly continuous), 3) the value of the attenuation of the evolved population of galactic origin modeled with the LMC law \citep{gordon2003a}, 4) the value of the attenuation of the younger population modeled with the starburst attenuation law \citep{calzetti2001a}, 5) the ratio $r$ of the mass of gas transformed into stars between the young burst and the evolved stellar population ranging from $\log r=-2.75$ to $\log r=2$. The total number of models considered for each star forming region is up to $127\times10^6$. When there is only one stellar population formed in the collision debris, there are only 3 free parameters.

\subsection{Mid-infrared spectral energy distribution modeling}

Coupled with dust models, mid- and far-infrared data place constraints on the dust (size distribution of the grains, ionization of the PAH, etc.). We carry out such a study in star forming regions in collision debris. To do so we compare with templates of  galaxies of various types and with dust models.

\subsubsection{Galaxy templates}

As a first step, we perform a simple comparison between the observed UV-to-mid-IR SED of the collision debris  and SED of template galaxies. Such a comparison is legitimate as  our regions have luminosities and sizes of typical dwarf galaxies \citep{boquien2009a}.  \citet{galliano2008a} modeled the SED of 35 nearby galaxies of different types, sizes, star formation activities, and ages with metallicities ranging from  $\simeq1/50$ to $\simeq2\;Z_\odot$. We scale each template to fit the UV-to-near-IR broadband observations and select the one which minimizes the $\chi^2$. This process allows us to compare the infrared SED of our sources to those of galaxies with similar stellar spectra.

\subsubsection{Dust modeling}

The intergalactic star-forming regions presented in this paper have been observed only in the mid-infrared and shortward. We have fluxes only up to 8~$\mu$m or 24~$\mu$m on a select number of systems. The absence of far-IR data makes the modeling of these objects extremely speculative. Indeed, the far-IR emission provides the most important and necessary constraints for the modeling of the dust SED of star forming region in collision debris. The far-IR emission by dust is dominated by grains in thermal equilibrium with the radiation field. Therefore, for a given set of grain optical properties, the far-IR SED gives direct access to the distribution of equilibrium temperatures of grains located mainly in dense PDR (photodissociation regions) and in the diffuse ISM (interstellar medium). Consequently, it is the most reliable tracer of the physical conditions that the bulk of the dust is exposed to. On the contrary, the mid-IR emission is dominated by grains stochastically heated. The extent of their temperature fluctuations is unrelated to the energy density of the general radiation field.

\begin{figure}[htbp]
\includegraphics[width=\columnwidth]{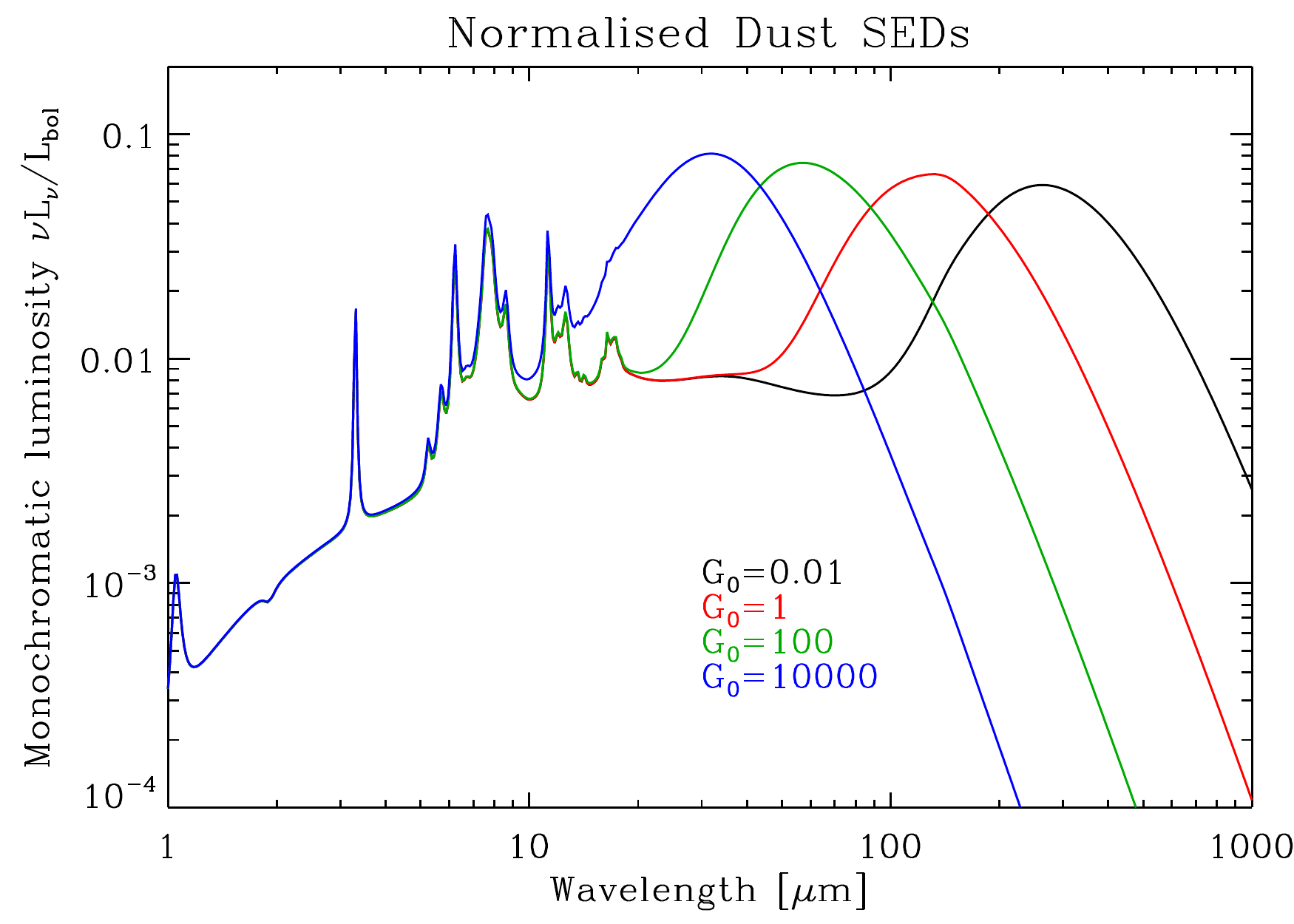}
\caption{Dust SED of grains exposed to ISRF (interstellar radiation field) of various intensities. The grain properties (abundances, size distributions and optical properties) are from the bare grain model with solar abundance constraints of \citet{zubko2004a}. The ISRF has the Galactic shape \citep{mathis1983a}, and is scaled by the factor $G_0$ ($G_0=1$ for the diffuse Galactic ISM). Each SED is normalized to its bolometric luminosity.}
\label{fig:sed_Zubko}
\end{figure}
Fig.~\ref{fig:sed_Zubko} illustrates this phenomenon. It shows the dust SED of the diffuse Galactic ISM exposed to ISRF of different intensities. The dust is made of carbon and silicate grains of different sizes (3.5~\AA\ to 0.25~$\mu$m) responsible for most of the continuum emission, and of PAH containing 10 to 1000 C atoms, mainly responsible for the mid-IR emission features. The integrated intensity of the ISRF is quantified by the standard $G_0$ parameter, defined by:
\begin{equation}
  G_0 = \frac{\displaystyle
              \int_{0.0912\;\mu m}^{0.2\;\mu m} 4\pi J_\lambda\, d\lambda}
             {1.6\times 10^{-6}\;\rm W\,m^{-2}},
\end{equation}
where $J_\lambda$ is the monochromatic mean intensity of the ISRF which is determined by the fit of the photospheric+nebular model from FUV to near-infrared. Using this ISRF, we compute the temperatures of our dust composition following the method of \cite{guhathakurta1989a}. Increasing the intensity of the ISRF makes the equilibrium grain SED shift to shorter wavelengths (because of hotter temperatures), while the spectrum of PAH and very small grains remains almost identical, since they are excited by single photon events. From Fig.~\ref{fig:sed_Zubko}, it is obvious that in the absence of far-IR observations, the mean intensity of the radiation field is unconstrained.

Unfortunately, we do not have any constraints in the far-infrared. Consequently we do not pretend to constrain the distribution of equilibrium grain temperatures within the galaxy. In addition, without further information, the diagnostics that we can extract from the mid-IR are very limited. However, we can put some constraints on the dust properties by relying on the information provided by the attenuation estimate (section~\ref{sssec:extinction}). Indeed, by assuming that the global emission from the star forming region is isotropic, the energy absorbed in the UV  and in the optical -- calculated from the attenuation determined by the photospheric+nebular fit -- should be re-radiated in the IR.

Another important point is that the ISRF illuminate the PAHs. Indeed, for a typical cloud, the radiation field becomes redder, when the optical depth increases. The shape of this radiation field will have an effect on the emissivity of the PAHs, which are stochastically heated, but will not have any effect on the equilibrium grains which are only sensitive to the total absorbed power. In our case, we consider the unattenuated ISRF (optically thin case), since PAH emission is dominated by photodissociation regions and cirrus clouds, which are both optically thin. This assumption introduces an additional uncertainty on the determination of the PAH--to--dust mass ratio that we will derive in the next paragraph. However, this error should be small compared to the variations we probe here as we will see in Sec.~\ref{sec:pahvsZ}.

Thus, we can perform a simple SED fit, where the total power emitted by the dust is determined by the energy balance, and where the PAH--to--dust mass ratio is constrained by the mid-IR fluxes. Indeed, in first approximation, $L_8\propto M_{PAH}\times G_0$. The fact that PAH are stochastically heated guarantees that they emit at the same wavelengths. Hence, only one point is needed to scale their emission. The total infrared power follows a similar relation: $L_{IR}\propto M_{dust}\times G_0$. Therefore the 8~$\mu$m-to-IR luminosity is proportional to the PAH-to-dust mass ratio: $L_8/L_{IR}\propto M_{PAH}/M_{dust}$. As shown in Fig.~\ref{fig:sed_Zubko}, the PAH-to-dust mass ratio, $f_{\mbox{\scriptsize PAH}}$, is approximately independent of $G_0$. Such an estimate of $f_{\mbox{\scriptsize PAH}}$ is strongly dependent on the accuracy of the attenuation estimate, since we use the attenuation to guess the level of the far--IR emission. It is therefore not extremely reliable, in itself. However, the comparison of this derived PAH--to--dust mass ratio to its expected value from chemical evolution considerations is more interesting, since it provides an insight on the amount of embedded IR emission. In other words, the attenuation estimate from the visible recombination lines is biased towards evolved star forming regions, where the molecular cloud covering factor is low enough for the radiation to escape. On the contrary, the attenuation towards very young embedded regions will not be traced by the Balmer decrement. In consequence, comparing the derived PAH-to-dust mass ratio to its expected value is a way to estimate the fraction of embedded infrared power. Another way to probe the presence of embedded star-forming regions would be to compare the estimate of the dust masses with the attenuation. Unfortunately, this would require to make assumptions on the geometry of the dust and the stellar populations, which is beyond the scope of this paper. This also prevents us from using the 24~$\mu$m emission to trace embedded star-forming regions as it depends on many unconstrained physical parameters such as the compacity of star-forming regions and the composition of the grains. The PAHs are a better tracer as they do not exhibit systematic temperature variations and finally we do not have 24~$\mu$m observastions for all systems.

\section{Results}
\label{sec:results}
Spectral energy distribution fits allow us, as we have shown earlier, to determine a number of physical parameters such as the age, the attenuation, the star formation history. We apply the modeling technique described earlier to our sample of star--forming collision debris. 

We present in Tab.~\ref{tab:fluxes} the list of fluxes, corrected for the foreground Galactic attenuation, we use in this paper. Some fluxes might very slightly differ from the ones published in \cite{boquien2009a} when the aperture has been modified.

The spectroscopically derived oxygen abundance, EW(H$\beta$) and $A_V$ are given in Table~\ref{tab:spec-parameters}.

\tabletypesize{\Tiny}
\begin{deluxetable*}{ccccccccccccccccccccccccccc}
\footnotesize
\tablecolumns{16}
\tablewidth{0pc}
\tablecaption{Fluxes}
\tablehead{
\colhead{System}&\colhead{FUV}&\colhead{NUV}&\colhead{B}&\colhead{V}&\colhead{R}&\colhead{J}&\colhead{H}&\colhead{K}&\colhead{N2}&\colhead{3.6}&\colhead{4.5}&\colhead{8.0}&\colhead{24}\\
&\colhead{$\mu$m}&\colhead{$\mu$m}&\colhead{$\mu$m}&\colhead{$\mu$m}&\colhead{$\mu$m}&\colhead{$\mu$m}&\colhead{$\mu$m}&\colhead{$\mu$m}&\colhead{$\mu$m}&\colhead{$\mu$m}&\colhead{$\mu$m}&\colhead{$\mu$m}&\colhead{$\mu$m}}\\
\startdata
Stephan's Quintet 2&$28.2\pm1.9$&$36.9\pm2.3$&$125.6\pm34.3$&$149.1\pm31.0$&$194.2\pm30.6$&--&--&--&--&$215.3\pm50.0$&$126.7\pm25.6$&$1192.4\pm87.9$&$888.6\pm113.0$\\
Stephan's Quintet 5&$5.5\pm1.4$&$8.8\pm1.4$&$12.3\pm2.5$&$15.4\pm1.2$&$23.9\pm1.3$&--&--&--&--&$39.1\pm12.8$&$32.5\pm4.0$&$766.5\pm52.1$&$1871.8\pm186.5$\\
Arp~105S&$72.0\pm3.7$&$134.8\pm8.1$&$251.3\pm23.4$&$292.8\pm37.5$&$376.7\pm82.1$&$286.2\pm82.1$&$265.6\pm202.8$&$289.9\pm111.8$&--&$489.3\pm191.3$&$347.9\pm128.6$&$2563.6\pm711.7$&--\\
Arp~245N&$83.7\pm3.8$&$154.1\pm2.5$&$1397.6\pm59.0$&$2224.2\pm38.2$&$2827.6\pm52.5$&$7686.6\pm2526.7$&$7287.1\pm3084.3$&$8212.1\pm4140.4$&$5872.4\pm880.9$&$3630.0\pm50.3$&$2200.0\pm98.2$&$9140.0\pm324.7$&--\\
NGC~5291N&$365.3\pm35.6$&$445.2\pm31.5$&$754.3\pm64.9$&$834.5\pm64.5$&$978.2\pm109.0$&$477.7\pm45.0$&$455.9\pm105.0$&$548.0\pm181.0$&--&$638.2\pm38.4$&$534.0\pm51.3$&$5235.6\pm89.6$&$15030.4\pm335$--\\
NGC~7252NW&$39.7\pm1.3$&$41.5\pm1.8$&$91.3\pm11.8$&$91.8\pm11.4$&$107.0\pm12.3$&--&--&--&$132.4\pm11.9$&$137.9\pm30.9$&$86.6\pm23.3$&$946.1\pm49.4$&$1259.8\pm289.7$\\\hline
System&FUV&NUV&u&B&g&V&r&R&i&z&3.6&4.5&8.0&24\\
VCC~2062&$47.2\pm4.4$&$64.9\pm6.0$&$163.0\pm114.7$&$234.5\pm39.8$&$251.3\pm47.6$&$259.4\pm27.1$&$268.6\pm63.0$&$281.4\pm38.7$&$253.3\pm74.2$&$271.7\pm215.0$&$210.0\pm52.0$&$185.0\pm55.0$&$1597.0\pm387.0$&$1188.1\pm876.2$
\enddata
\label{tab:fluxes}
\end{deluxetable*}

\tabletypesize{\normalsize}
\begin{deluxetable*}{cccccccccc}
\tablecolumns{16}
\tablewidth{0pc}
\tablecaption{Spectroscopic characteristics}
\tablehead{
\colhead{System}&\colhead{Oxygen abundance}&\colhead{EW(H$\beta$)}&\colhead{$A_V$}\\
\colhead{}&\colhead{$12+\log O/H$}&\colhead{\AA}}&\colhead{}\\
\startdata
Stephan's Quintet 2&8.8&--&--\\
Stephan's Quintet 5$^a$&8.7&137--454&1.3--2.3\\
Arp~105S$^b$&8.4&111&0.8\\
Arp~245N$^c$&8.6--8.7&6--13&2.0\\
NGC~5291N$^d$&8.4&140&$<0.1$\\
NGC~7252NW$^e$&8.6&30&1.1\\
VCC~2062$^f$&8.6--8.7&70&0.1
\enddata
\tablecomments{a=Duc et al. (in preparation), b=\cite{duc1994a}, c=\cite{duc2000a}, d=\cite{duc1998a}, e=\cite{duc1995a}, f=\cite{duc2007a}.}

\label{tab:spec-parameters}
\end{deluxetable*}

The best photospheric and nebular fit parameters are given in Table~\ref{tab:fit-parameters}. Information obtained from the dust models is presented in Table \ref{tab:pahvsZ}.

\subsection{Stephan's Quintet}

Stephan's Quintet shows signs of interaction \citep{moles1997a} with prominent intergalactic star-forming regions associated and large reservoirs of atomic and molecular hydrogen \citep{yun1997a,lisenfeld2002a,williams2002a,appleton2006a,cluver2010a} but also hot stripped gas \citep{sulentic1995a}. We have decided to study two regions of particular interest. The famous and prominent objects called SQ-A and SQ-B are not studied here: their high optical attenuation make any fit of their SED very uncertain. See Fig. 5 in \cite{boquien2009a} for an image of the system and the selected regions:
\begin{itemize}
\item The so-called ``tip-tail'' region (region 2 in Figure~5 in \cite{boquien2009a} and hereafter) is located at the tip of the stellar tidal tail that emanates from NGC 7319. Its optical morphology and rather red color reveal the likely presence of dust lanes, and of an important old stellar component likely tidally expelled from the nearby spiral. Its vicinity to SQ-B in the tidal tails suggest that they have close chemical properties.
\item Region 5, on the contrary, appears as a very blue and compact star forming complex located towards a gaseous tidal tail. Optical images do not show any evidence of an evolved stellar population which makes it particularly interesting.
\end{itemize}

\subsubsection{Stellar populations}

The near-infrared band observations that are available in the archives are unfortunately too shallow to detect the selected star forming regions. However, given the apparent importance on optical images of the old stellar population in region 2, we have initially set the ratio $r$, of the mass stars produced by the second burst divided by the mass of the stars produced by the first burst, to the rather high value of $\log r=-1$.

The free parameters are the star formation history and the attenuation of both populations. 

\begin{figure*}[!htbp]
\includegraphics[width=\textwidth]{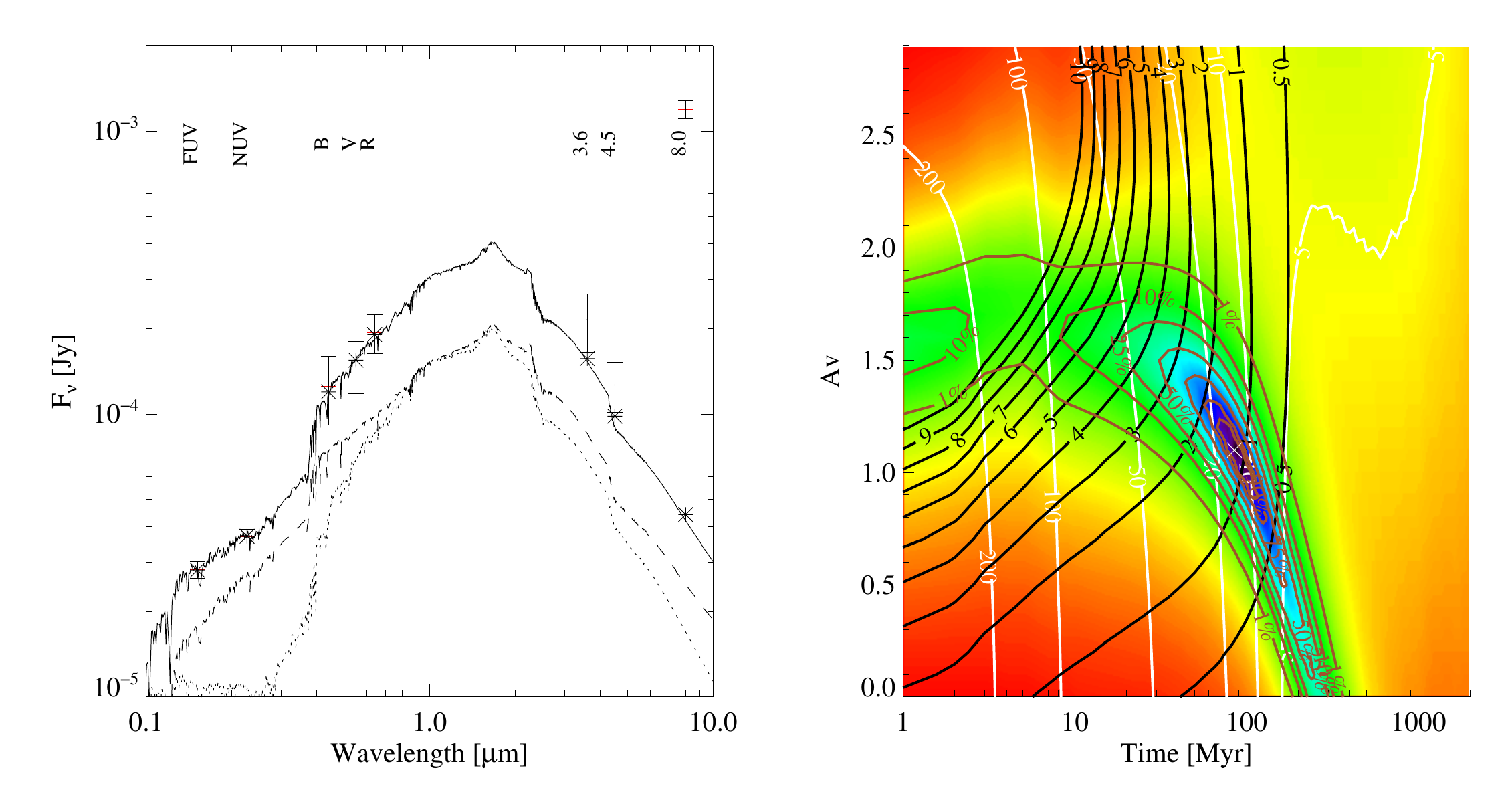}
\caption{Modeling of region 2 in Stephan's Quintet. We use a mass ratio of young to old stars $\log r=-1$. The fit yields no attenuation for the evolved population of galactic origin, and a timescale $\tau=37\times10^6$ years. The fit has been performed taking into account the FUV, NUV, B, V and R-band data. Given their uncertainties, the H$\alpha$ flux and the H$\beta$ equivalent width have not been used to constrain the fits, but are included in the plot as an indication. Left: the spectral energy distribution of the best fit. The dotted line represents the spectral energy distribution of the population originating from the parent galaxies, the dashed line represents the population formed by the collision debris and the solid line represents the sum of both populations. The star symbols indicate the flux in the filters when convolved with the spectral energy distribution, taking into account the nebular emissions, whereas the red lines crossing the error bars represent the observations. Right: the $\chi^2$ as a function of the age and the attenuation of the young stellar component. The dark blue color indicates low $\chi^2$ values and red high ones. The white contours indicate the value of the H$\beta$ equivalent width in \AA, and the black ones the ratio of the H$\alpha$ from the model to the observed value. The attenuation is taken into account. Finally the brown contours follow the $\chi^2$ color scheme and give the probability that given model parameters can reproduce the observations. The cross shows which model SED has the highest probability to reproduce the observations.\label{fig:fit-SQ-2}}
\end{figure*}

As shown in Figure~\ref{fig:fit-SQ-2}, the best fit (able to reproduce the observations with a probability of 98\%) is obtained with an age of $t=85\times10^6$ years, no attenuation for the evolved population of galactic origin, an attenuation of 1.1 magnitudes for the one formed in the intergalactic medium and a time constant $\tau=37\times10^6$ years. The H$\alpha$ flux is also reproduced relatively well. The range of possible attenuations however is large: from 0.7 to 1.4 magnitudes.

Assuming an attenuation of 1.1 magnitude,  all time constants are possible with a probability of at least 75\% to reproduce the observations. For a short timescale, typically $\tau<10\times10^6$ years, the age is about $t\simeq30\times10^6$ years. Conversely a nearly continuous star formation yields an age comprised between $t=100\times10^6$ years and $t=200\times10^6$ years.

We have to note that the presence of a larger evolved population would redden the optical spectrum, which as a consequence would make the model-derived age of the population formed in the intergalactic medium younger. The ages we find should then rather be seen as upper limits. In case there were proportionally fewer evolved stars, the ages would be older.

For the blue -- almost pristine -- region 5, we have set $\log r=2$.

\begin{figure*}[!htbp]
\includegraphics[width=\textwidth]{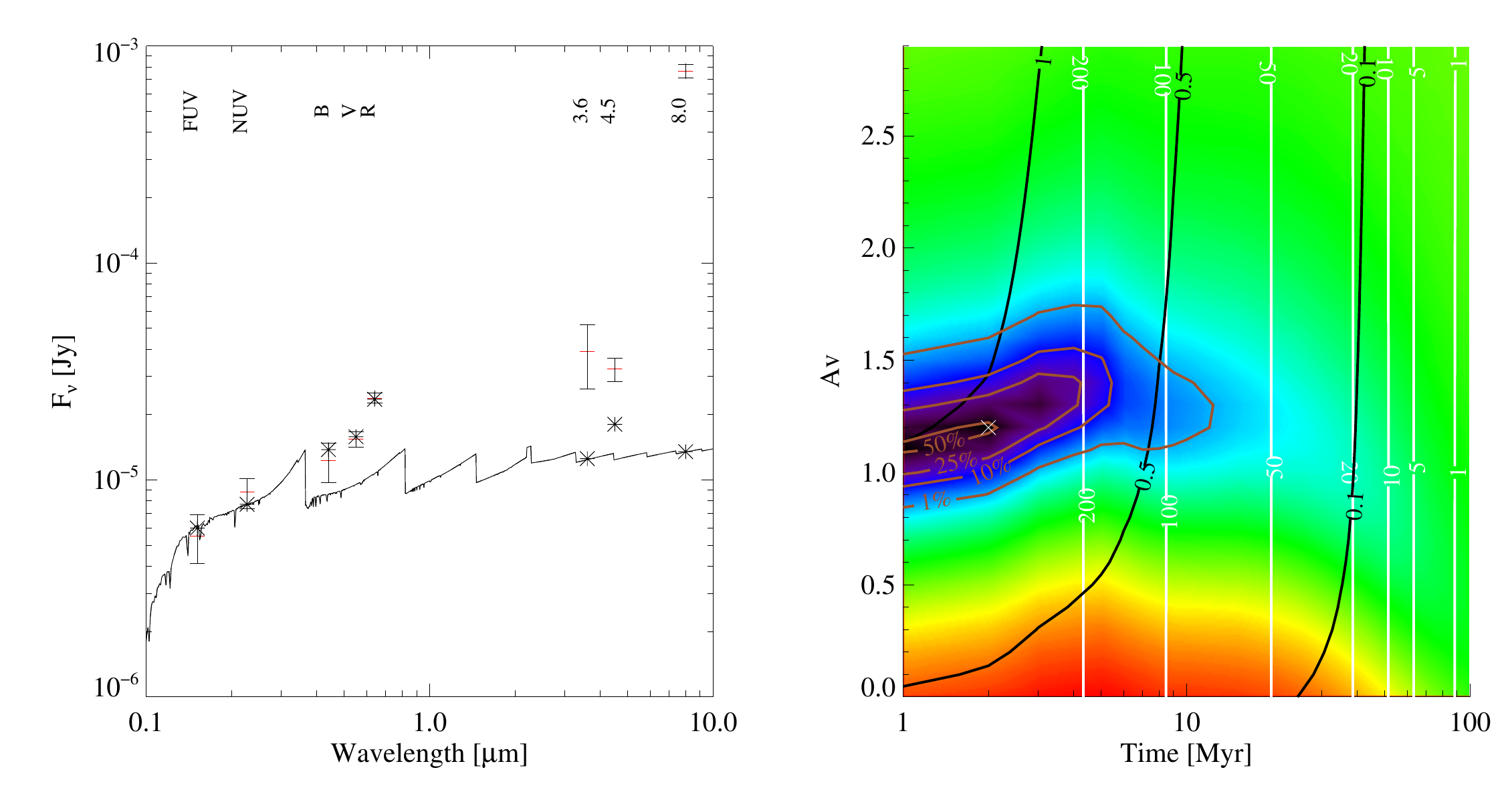}
\caption{Modeling of region 5 of the Stephan's Quintet with a mass ratio $\log r=2$, and a timescale $\tau=13\times10^9$ years. The fit has been performed on the FUV, NUV, B, V and R bands. See Figure~\ref{fig:fit-SQ-2} for additional details.\label{fig:fit-SQ-5}}
\end{figure*}

The fit in Figure~\ref{fig:fit-SQ-5} shows that unsurprisingly the burst of star formation is young, typically less than $t=10\times10^6$ years with the attenuation comprised between 1.1 and 1.2 magnitudes, relatively close to what has been measured spectroscopically, for a probability to reproduce the observations over 50\%. The best fit (reproducing the observations with a probability of 55\%) corresponds to an age of $t=2\times10^6$ years and an attenuation of 1.2 magnitudes. A burst age of a few tens of million years cannot reproduce the observations in any case. The age being young, the influence of the timescale is very marginal. 

\subsubsection{Dust}

\begin{figure*}[htbp]
\includegraphics[width=\columnwidth]{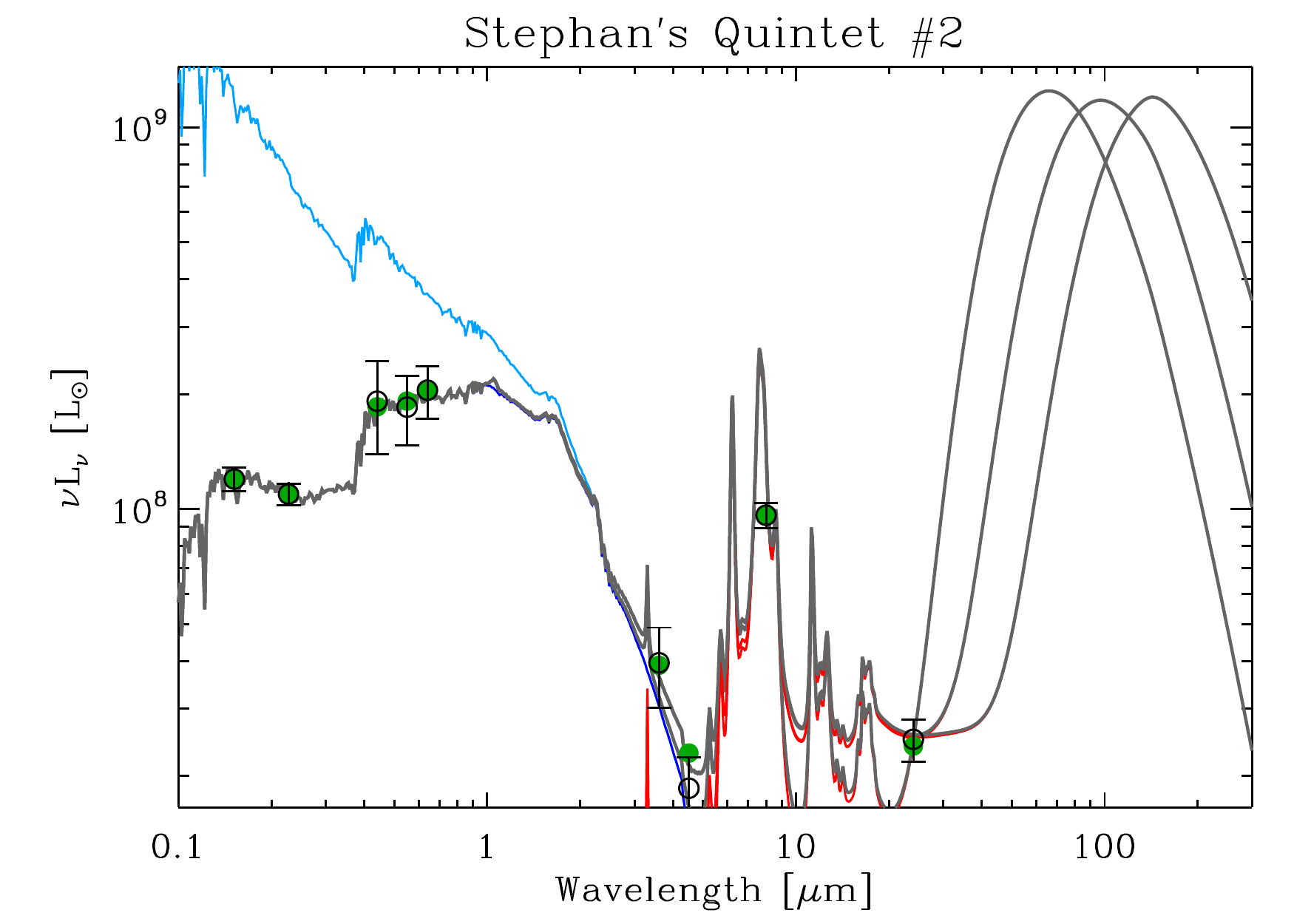}
\includegraphics[width=\columnwidth]{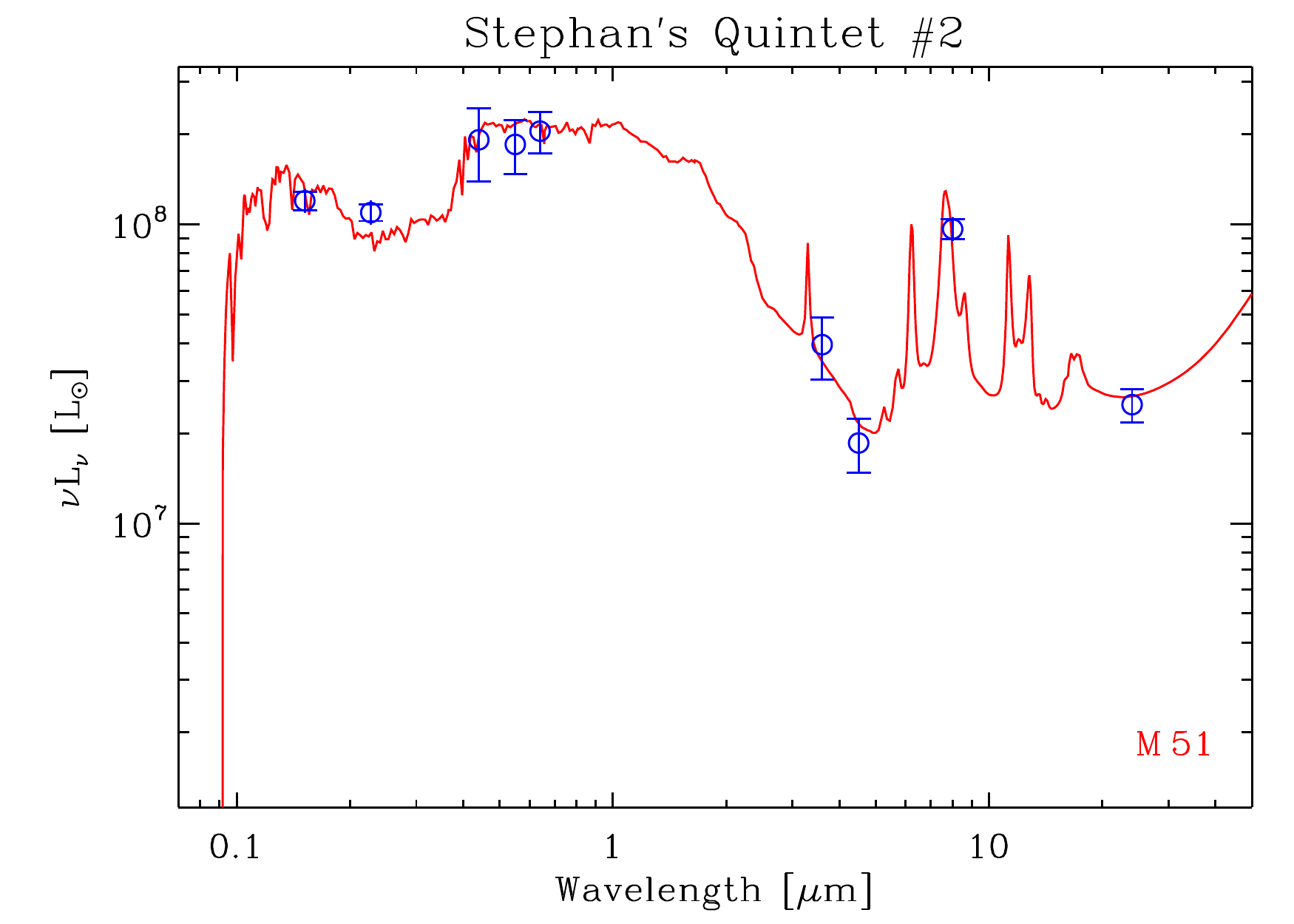}
\caption{Left: Fitted SED of region 2 of Stephan's Quintet. {\it Red:} dust emission; {\it blue:} stellar spectrum without attenuation; {\it cobalt:} attenuated stellar spectrum; {\it gray:} total; {\it green dots:} model integrated within the broadbands. The different curves correspond to different values of $G_0$: 1, 10 and 100. Right: Observed SED of region 2 of Stephan's Quintet (blue circles with error bars), compared to SED templates of nearby galaxies modeled by \citet{galliano2008a} (red line).\label{fig:dust-SQ2}}
\end{figure*}

In Figure \ref{fig:dust-SQ2} we present the fit of the SED of region 2 with its best template and  dust model.

The closest template to the observations is   M51, a metal-rich relatively dusty spiral galaxy. Although essentially constrained in the UV-to-optical regime, the match is very good up to the mid-IR. The lack of observations beyond 24~$\mu$m prevents us from constraining $G_0$.

\begin{figure*}[htbp]
\includegraphics[width=\columnwidth]{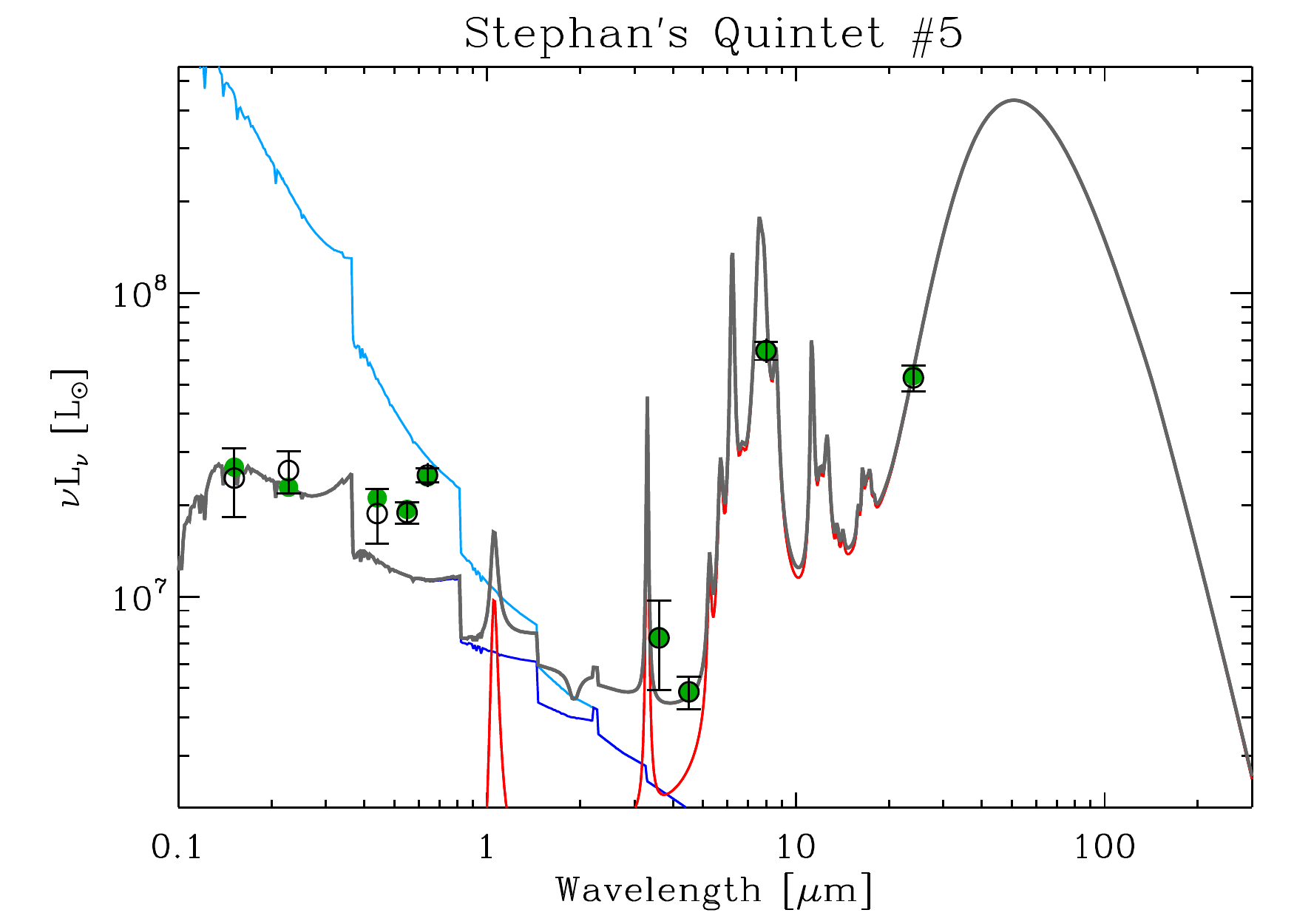}
\includegraphics[width=\columnwidth]{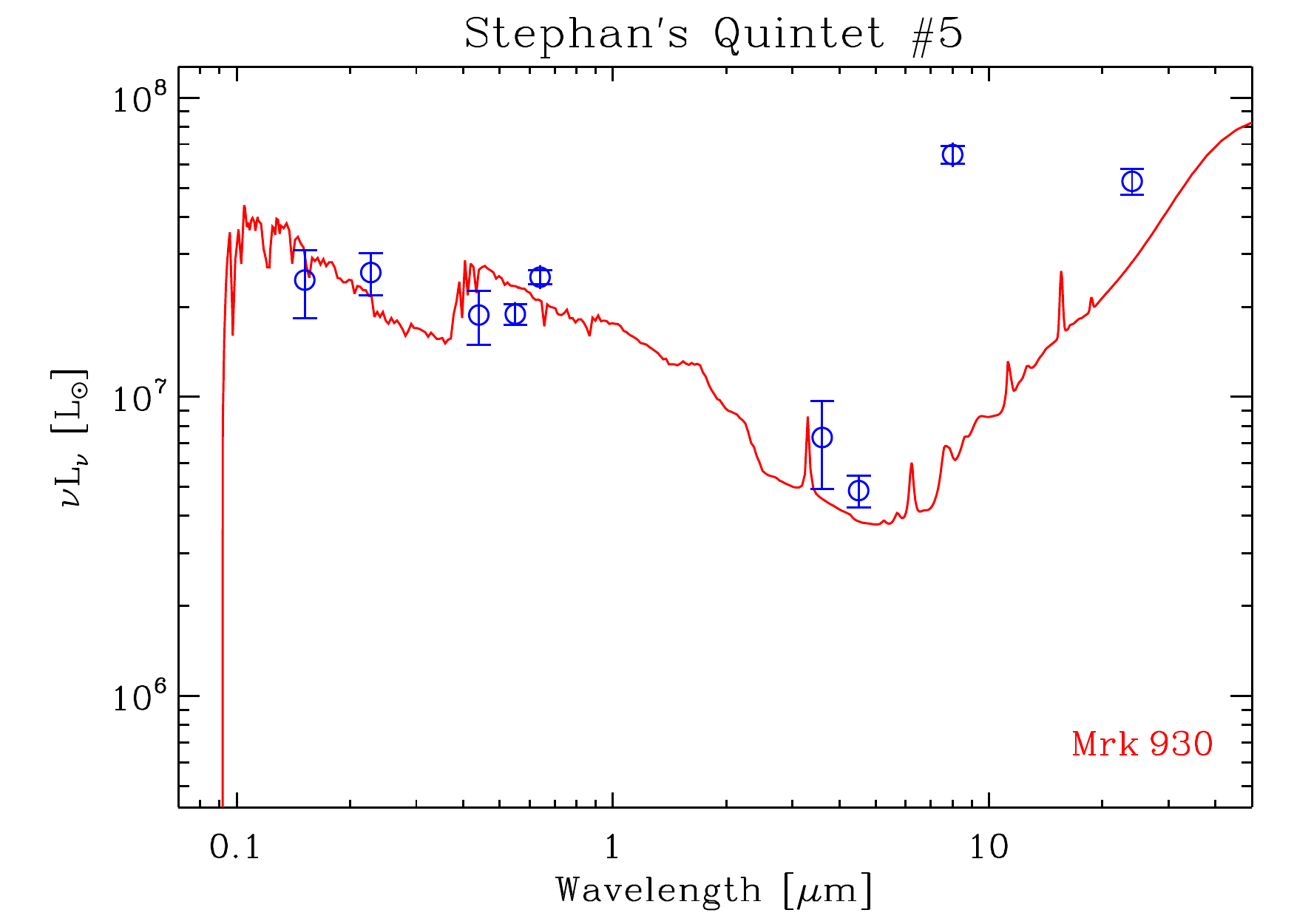}
\caption{Same as Figure \ref{fig:dust-SQ2} for Stephan's Quintet region 5. The value of $G_0$ is 520.\label{fig:dust-SQ5}}
\end{figure*}

The fit of the SED of region 5, with its best template and dust model is presented in  Figure \ref{fig:dust-SQ5}.

The best template in the UV-optical regime corresponds to Mrk~930, a low metallicity starbursting galaxy. In the mid-IR however, the fluxes in the intergalactic region are much higher, indicating important dust emission (consistent with its high metallicity). For the dust model, we show only one fit for a value of $G_0$ that was free to vary. The reason is that the 24~$\mu$m flux provides a constraint that cannot be explained with a simple very small grain continuum, without altering their size distribution. This fit is purely speculative, and we do not pretend that the far-IR SED will look as plotted. This fit simply shows the highest value of $G_0$ (here $G_0\simeq600$) dominating the IR SED. It is equivalent to the parameter $U_{\rm max}$ in the \citet{dale2001a} prescription.

\subsection{Arp 105S}
\label{ssec:arp105}
Arp 105S has the color and morphology of a blue compact dwarf galaxy. \cite{duc1994a} have shown that it is in fact a Tidal Dwarf Galaxy exhibiting prominent star--forming regions. The measured H$\beta$ equivalent width of 111~\AA\ suggests a fairly young burst.

Arp 105S is located towards the stellar halo of an elliptical galaxy \citep[Fig. 1 of][]{boquien2009a}. So as to minimize the pollution by this galaxy, we have subtracted its profile using the {\sc ellipse} and {\sc bmodel} procedures in {\sc iraf}. 

\subsubsection{Stellar populations}

\begin{figure*}[!htbp]
\includegraphics[width=\textwidth]{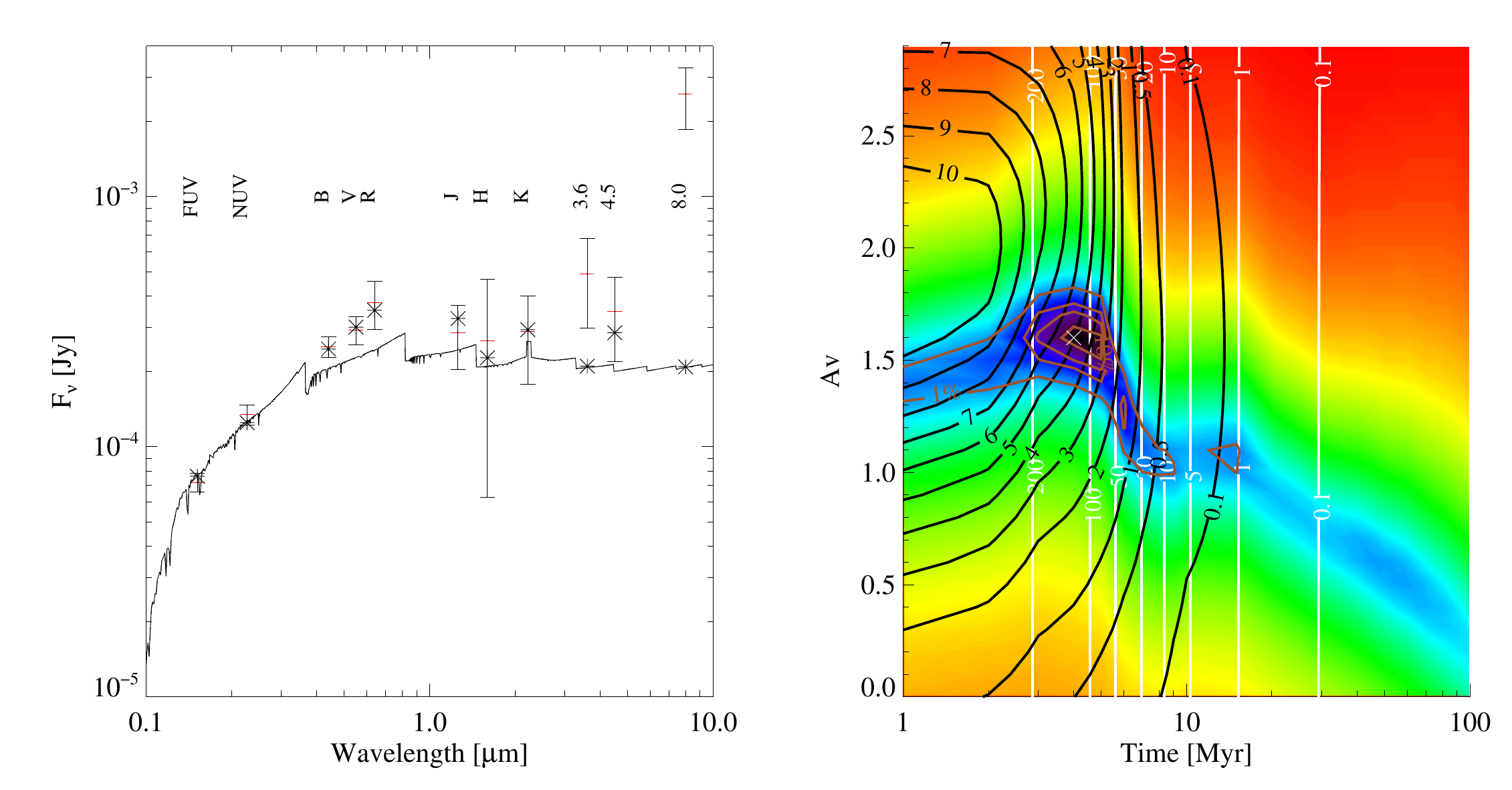}
\caption{Modeling of Arp~105S with a mass ratio $\log r=2$, and a timescale $\tau=10^6$ years. The fit has been performed on the FUV, NUV, B, V, R, J, H and K bands. See Figure~\ref{fig:fit-SQ-2} for additional details.\label{fig:fit-Arp-105S}}
\end{figure*}

The best fit (presented in Figure~\ref{fig:fit-Arp-105S}, with a probability of 67\% to reproduce the observations) corresponds to a burst strength of $\log r=2$, an age of $t=4\times10^6$ years, a very short time scale of $\tau=10^6$ years (i.e. an instantaneous burst) and an attenuation of 1.6 magnitudes for the population formed in the collision debris. The H$\alpha$ flux is however overestimated and the attenuation is larger than the one determined by \cite{duc1994a}. The ultraviolet flux is not perfectly adjusted by the model, which may be due to a slight remaining contamination by the elliptical galaxy.

\begin{figure*}[!htbp]
\includegraphics[width=\textwidth]{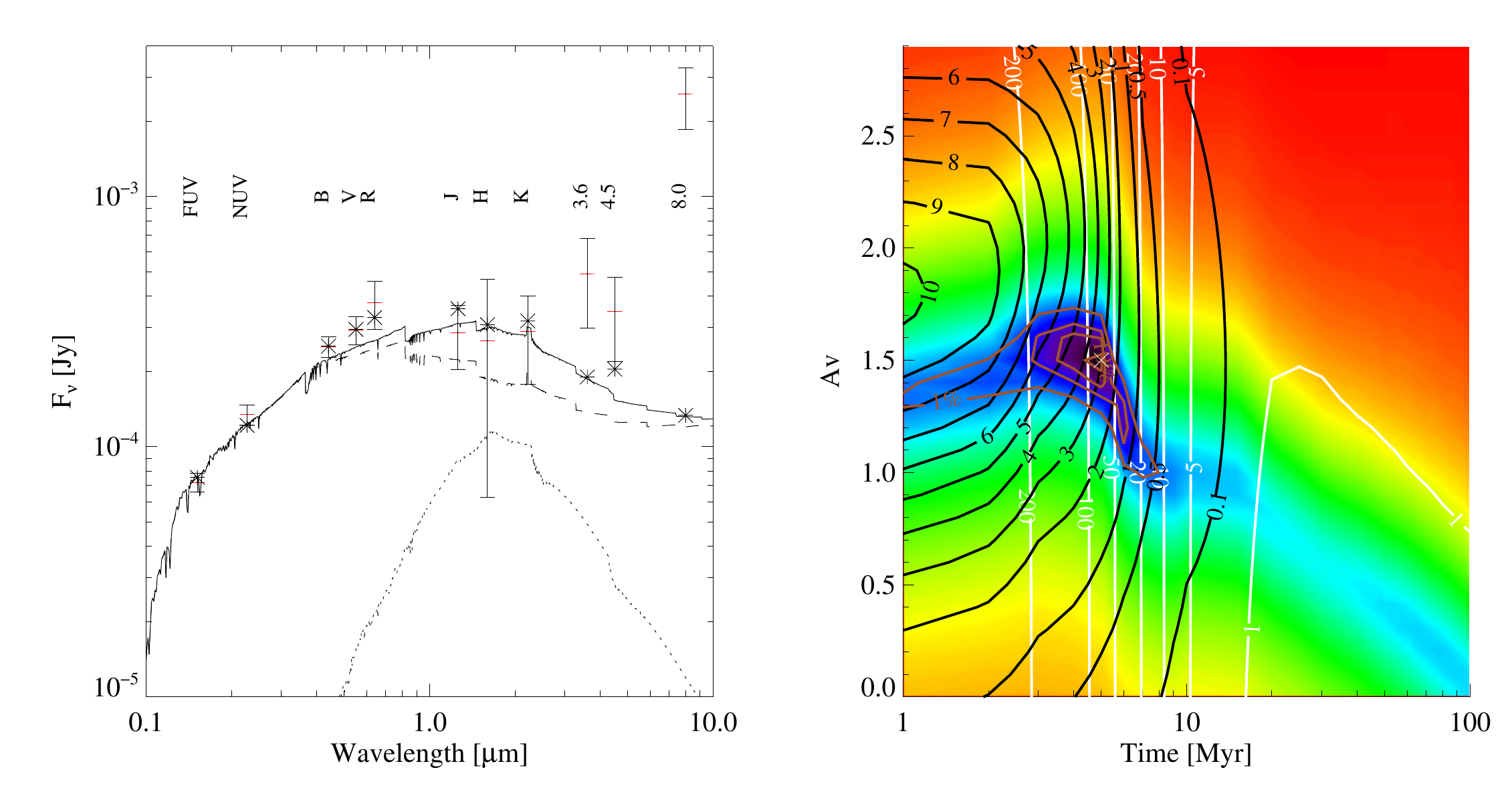}
\caption{Modeling of Arp~105S with a mass ratio $\log r=-1$, an attenuation of 1.4 magnitudes for the evolved component and 1.5 magnitudes for the population formed in collision debris and a timescale $\tau=10^6$ years. The fit has been performed on the FUV, NUV, B, V, R, J, H and K bands. See Figure~\ref{fig:fit-SQ-2} for additional details.\label{fig:fit-Arp-105S-r}}
\end{figure*}

To test whether the presence of an old population is compatible with the observations, we have also fitted the SED setting $\log r=-1$. The best fit (shown in Figure \ref{fig:fit-Arp-105S-r}) then corresponds to a probability that the model can reproduce the observations of 63\%.

The parameters are close to the ones obtained for $\log r=2$: $t=4\times10^6$ years, $\tau=10^6$ years and a slightly lower attenuation. Values well below $\log r=-1$ are excluded by our models.

\subsubsection{Dust}
\begin{figure*}[htbp]
\includegraphics[width=\columnwidth]{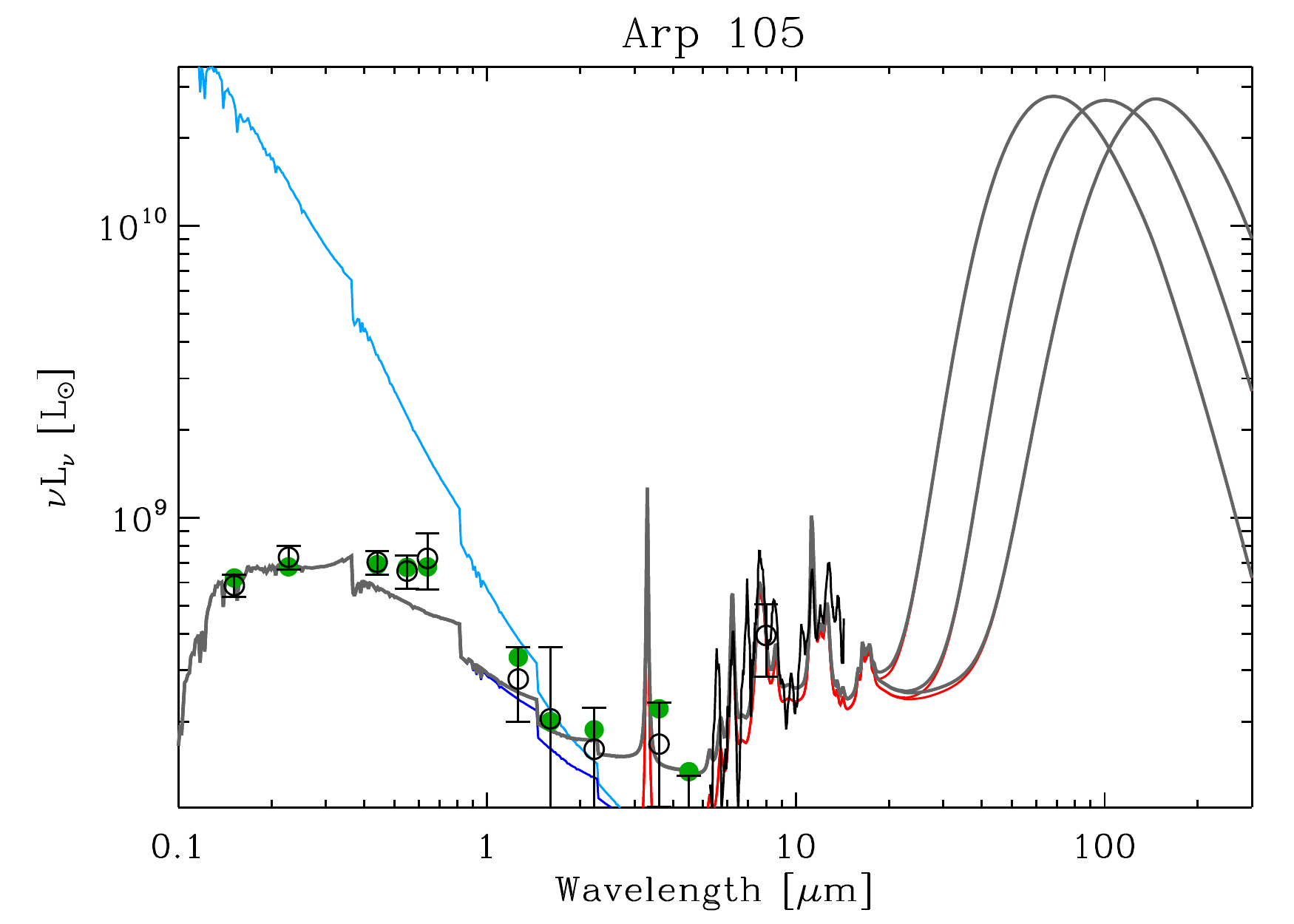}
\includegraphics[width=\columnwidth]{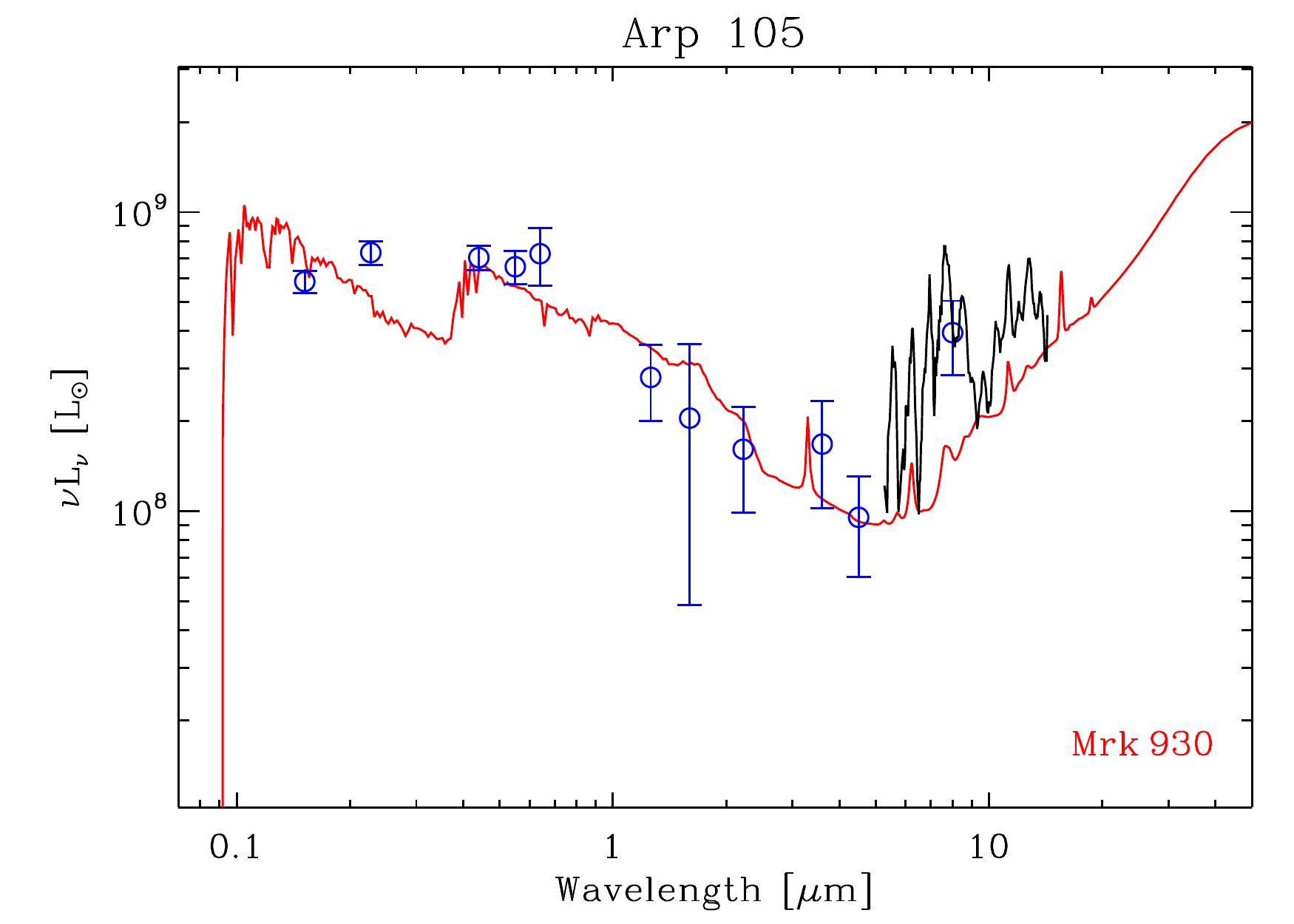}
\caption{Same as Figure \ref{fig:dust-SQ2} for Arp~105S. The black line represents the Spitzer IRS spectrum rescaled to fit the model \citep{boquien2009a}. The values of $G_0$ are 1, 10 and 100.\label{fig:dust-arp105}} 
\end{figure*}

In Figure \ref{fig:dust-arp105} we present the fit of Arp~105S with its best template and a dust model.

The best template for this blue star-forming region, similarly to the region 5 of Stephan's Quintet, is the starburst BCD Mrk~930. The mismatch in the mid-IR tells about a relatively high dust content.

The rescaled IRS spectrum allows us to accurately constrain the PAH-to-dust mass ratio. However, the fit of the spectrum is not perfect. In particular, the 6.2 to 7.7~$\mu$m feature ratio is lower than the modeled one. This peculiarity could be explained by a destruction of the smallest PAH in this particularly active star-forming object \citep[Fig.~14 of][]{galliano2008b}.

\subsection{Arp 245N}
As shown by \cite{duc2000a}, Arp~245N is a Tidal Dwarf Galaxy candidate located at the tip of a prominent stellar tidal tail \citep[Fig. 2 of][]{boquien2009a}. The optical images of Arp~245N show without any possible doubt the presence of an evolved population originating from the parent galaxies. The TDG exhibits several bluer regions, with H$\alpha$, UV and mid--IR counterparts, indicating the presence of on-going star formation. 

\subsubsection{Stellar populations}
\begin{figure*}[!htbp]
\includegraphics[width=\textwidth]{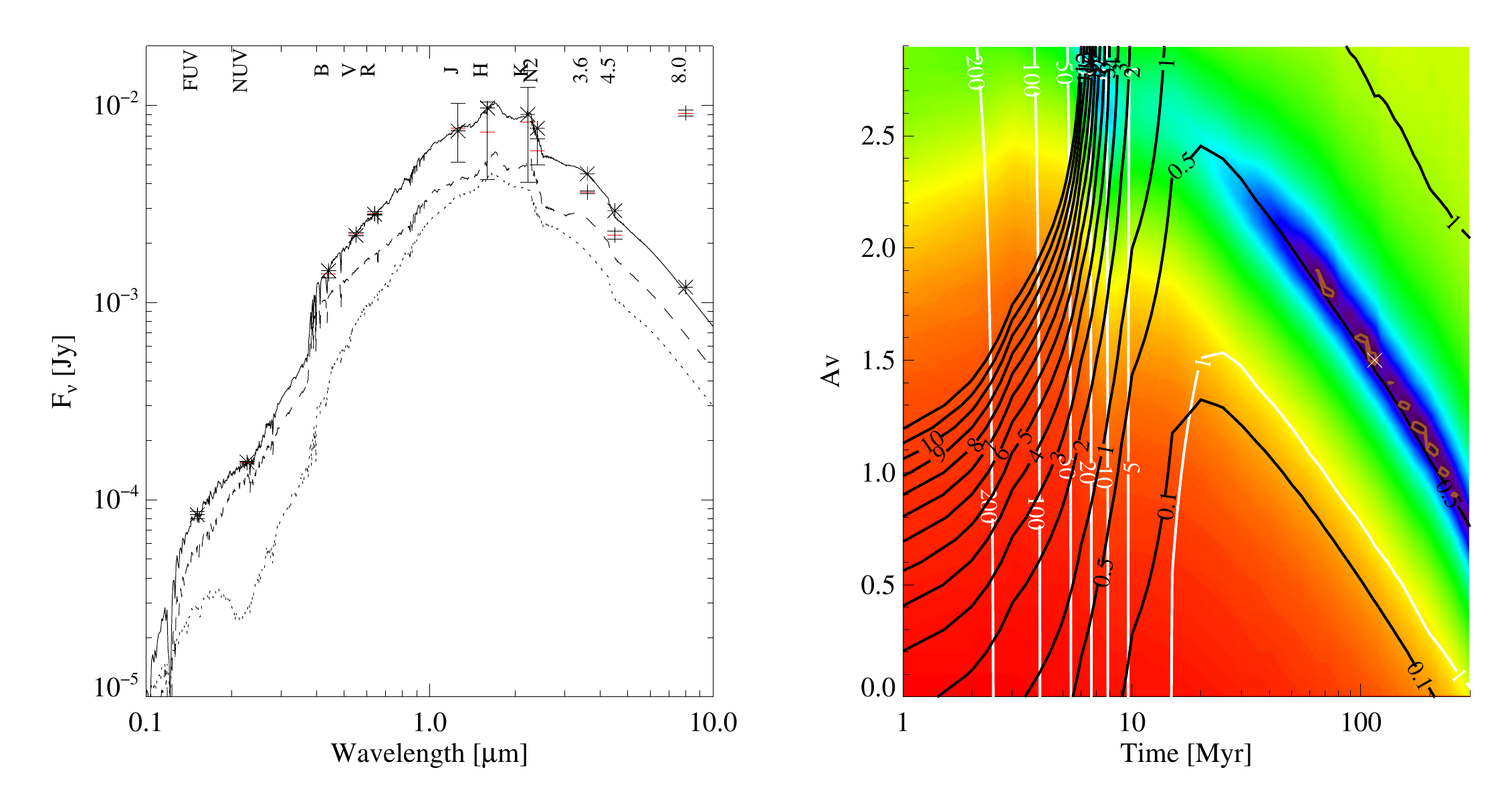}
\caption{Modeling of Arp~245N with a mass ratio $\log r=-0.75$, an attenuation of 1.1 magnitudes for the evolved component and a timescale $\tau=1\times10^6$ years. The fit has been performed on the FUV, NUV, B, V, R, J, H and K bands. See Figure~\ref{fig:fit-SQ-2} for additional details.\label{fig:fit-Arp-245N}}
\end{figure*}

The best fit (shown in Figure~\ref{fig:fit-Arp-245N}) corresponds to a burst strength of only $\log r=-0.75$, $t=115\times10^6$ years, a short timescale of $\tau=10^6$ years, an attenuation of 1.1 magnitude for the evolved population and 1.5 magnitudes for the population formed in the collision debris.

Only a narrow range of parameters has a non negligible probability to reproduce the observations. 

\begin{figure*}[!htbp]
\includegraphics[width=\textwidth]{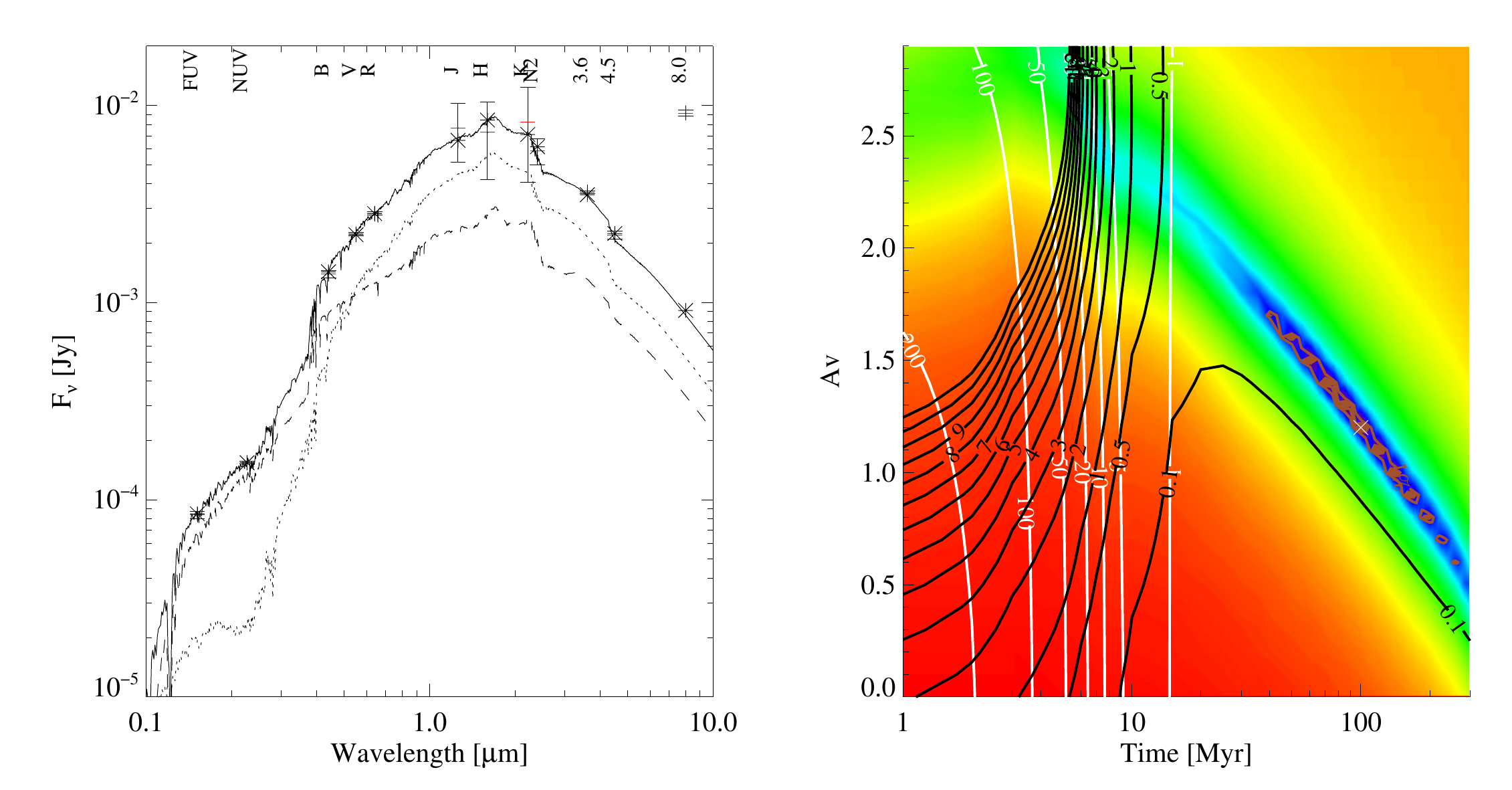}
\caption{Modeling of Arp~245N with a \cite{scalo1986a} initial mass function, a mass ratio $\log r=-0.75$, an attenuation of 0.6 magnitudes for the evolved component and a timescale $\tau=1\times10^6$ years. The fit has been performed on the FUV, NUV, B, V, R, J, H and K bands. See Figure~\ref{fig:fit-SQ-2} for additional details.\label{fig:fit-Arp-245N-scalo}}
\end{figure*}

Given the strong constraints given by the fit, we could for this system analyze the effect of changing the initial mass function. The best fit using a \cite{scalo1986a} IMF is shown in figure \ref{fig:fit-Arp-245N-scalo}.

The burst strength remains unchanged $\log r=-0.75$ and the other parameters change slightly: $t=100\times10^6$ years; $\tau=10^6$ years; attenuations of 0.6 magnitude and 1.2 magnitudes for the old and young stellar populations.

The small offset towards younger ages is due to the fact that the \cite{scalo1986a} IMF creates less massive stars, which have the strongest effect in the ultraviolet.

We note that with the \cite{scalo1986a} initial mass function, the IRAC bands at 3.6~$\mu$m and 4.5~$\mu$m are better reproduced than with the \cite{salpeter1955a} initial mass function. With the \cite{scalo1986a} IMF more intermediate and less very high or very low mass stars are formed than with the \cite{salpeter1955a} IMF which leads to lower 3.6/optical and 4.5/optical ratio.

The spectral energy distribution is best fitted assuming a \cite{scalo1986a} initial mass function. If we assume that it is the correct initial mass function, it means that the 3.6~$\mu$m and 4.5~$\mu$m bands are largely dominated by the stellar flux and contain very little pollution by a dust component. Unsurprisingly, \cite{boquien2009a} also found that the very small grains emission was small in Arp~245N compared to the PAH emission.

\subsubsection{Dust}
\begin{figure*}[htbp]
\includegraphics[width=\columnwidth]{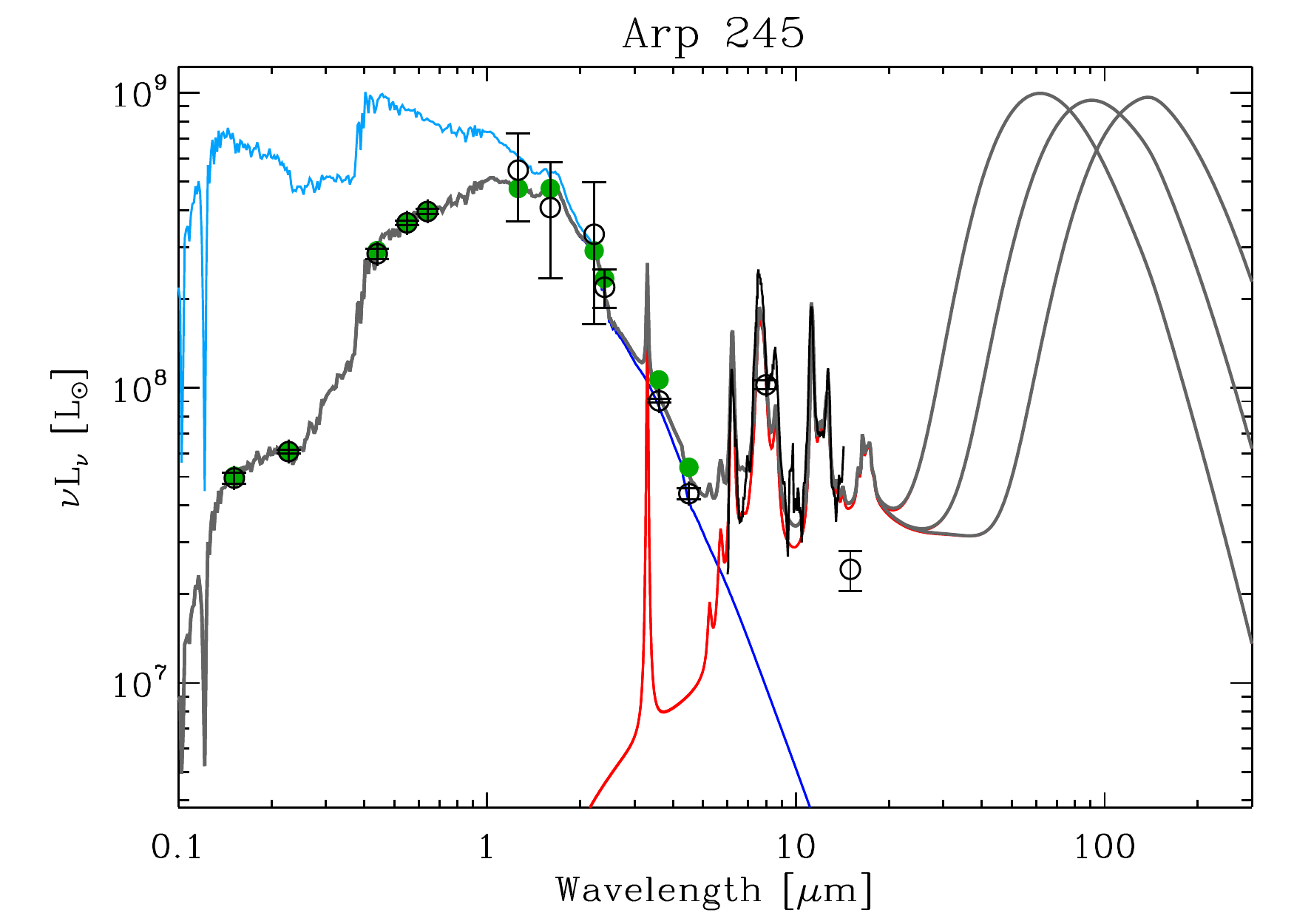}
\includegraphics[width=\columnwidth]{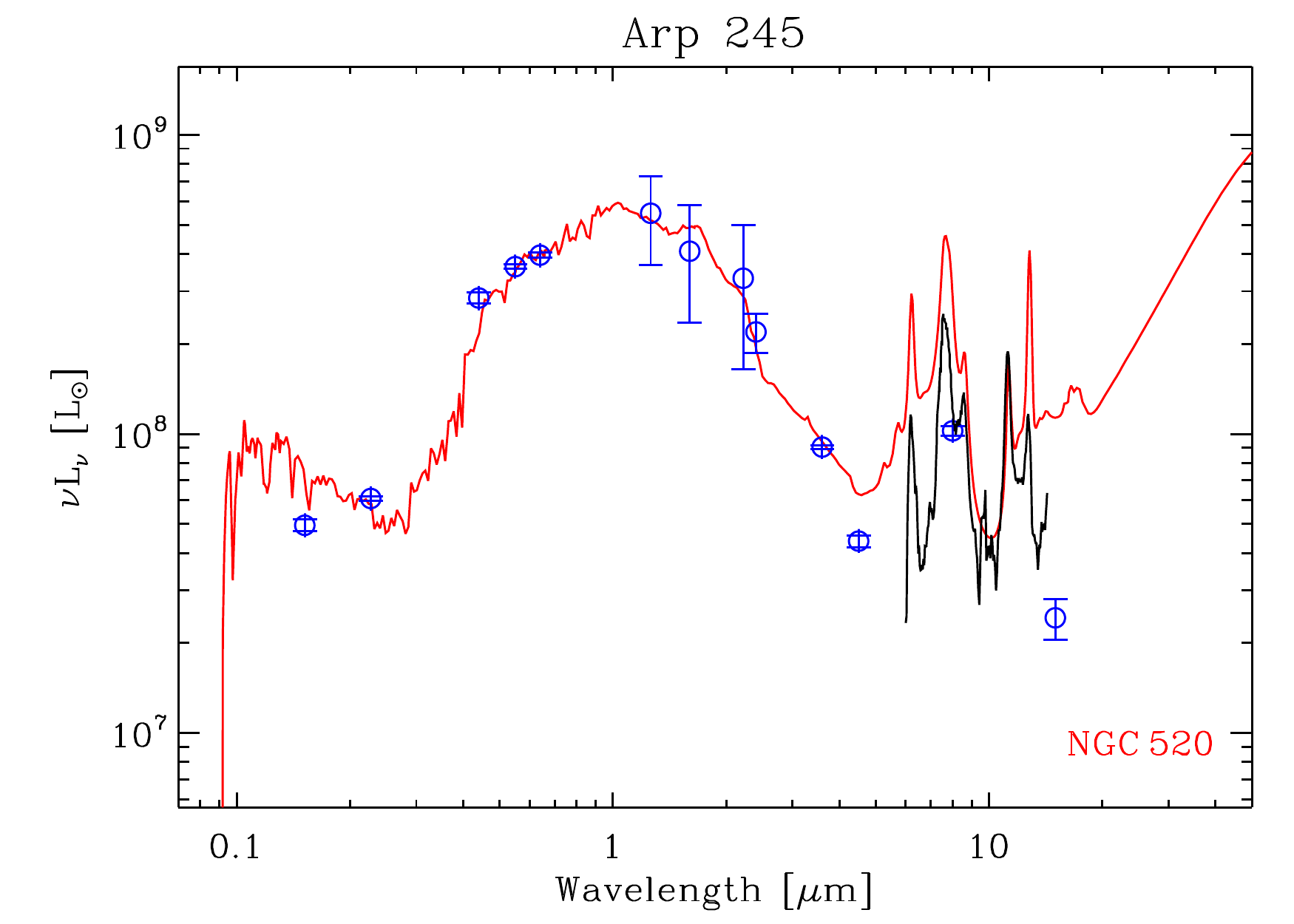}
\caption{Same as Figure \ref{fig:dust-SQ2} for Arp~245N. The black line represents the Spitzer rescaled IRS spectrum \citep{boquien2009a}. The values of $G_0$ are 1, 10 and 100.\label{fig:dust-arp245}} 
\end{figure*}

In Figure \ref{fig:dust-arp245} we present the fit of Arp~245N with its best template and a dust model.

The best template in the UV-to-IR regime corresponds to NGC~520, a starburst merger with composite stellar populations, which is consistent with our modeling of its stars. Interestingly, Arp~245N shows  weaker 6.2 to 7.7~$\mu$m and PAH emission compared to the one of NGC~520.

\subsection{NGC 5291N}

NGC~5291N is the most luminous star forming region in the collisional ring around NGC~5291 \citep[Fig. 3 of][]{boquien2009a}. Previous works by \cite{duc1998a,higdon2006a,boquien2007a} suggest that the star forming regions in this system are probably very young and dynamical arguments hint that there might not be any underlying evolved population. The nebular lines are particularly strong in this system as has been shown by \cite{duc1998a}, which makes a very accurate SED modeling more difficult. 

\subsubsection{Stellar populations}

\begin{figure*}[!htbp]
\includegraphics[width=\textwidth]{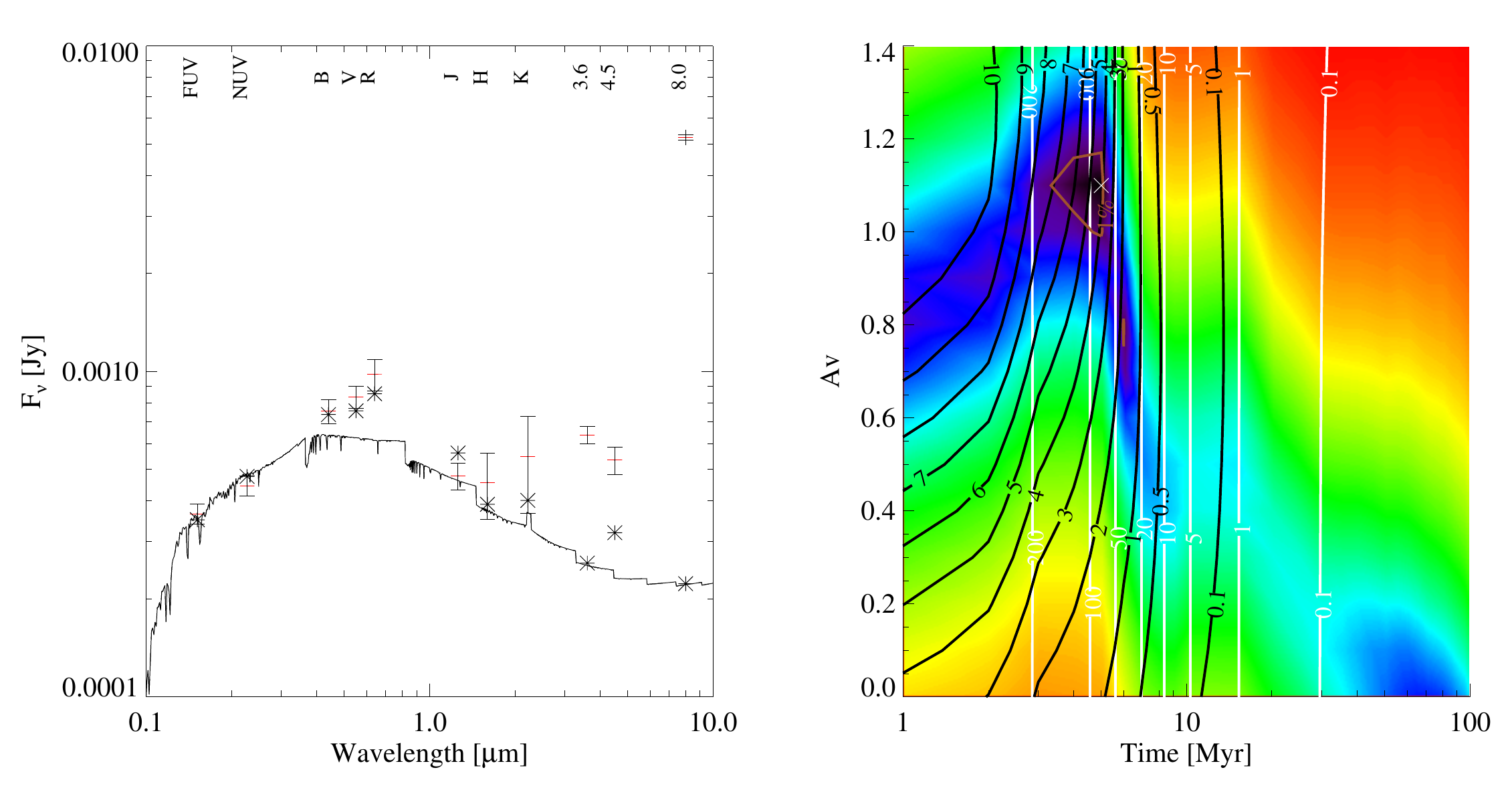}
\caption{Modeling of NGC~5291N with a mass ratio $\log r=2$, an attenuation of 1.1 magnitudes for the population formed in collision debris and a timescale $\tau=1\times10^6$ years. The fit has been performed on the FUV, NUV, B, V, R, J, H and K bands. See Figure~\ref{fig:fit-SQ-2} for additional details.\label{fig:fit-NGC5291N}}
\end{figure*}

The best fit (shown in Figure~\ref{fig:fit-NGC5291N}) indicates a rather young burst, $t=5\times10^6$ years with a short timescale $\tau=10^6$ years, an attenuation of 1.1 magnitudes and a high burst strength $\log r=2$. The flux in the J band is slightly overevaluated by the model. This may be due to the uncertainties on the emission line fluxes. For a young starburst this band is dominated by the Pa-$\beta$ line at 1.28181~$\mu$m and by the [Ne~II] line at 1.28~$\mu$m.

\begin{figure*}[!htbp]
\includegraphics[width=\textwidth]{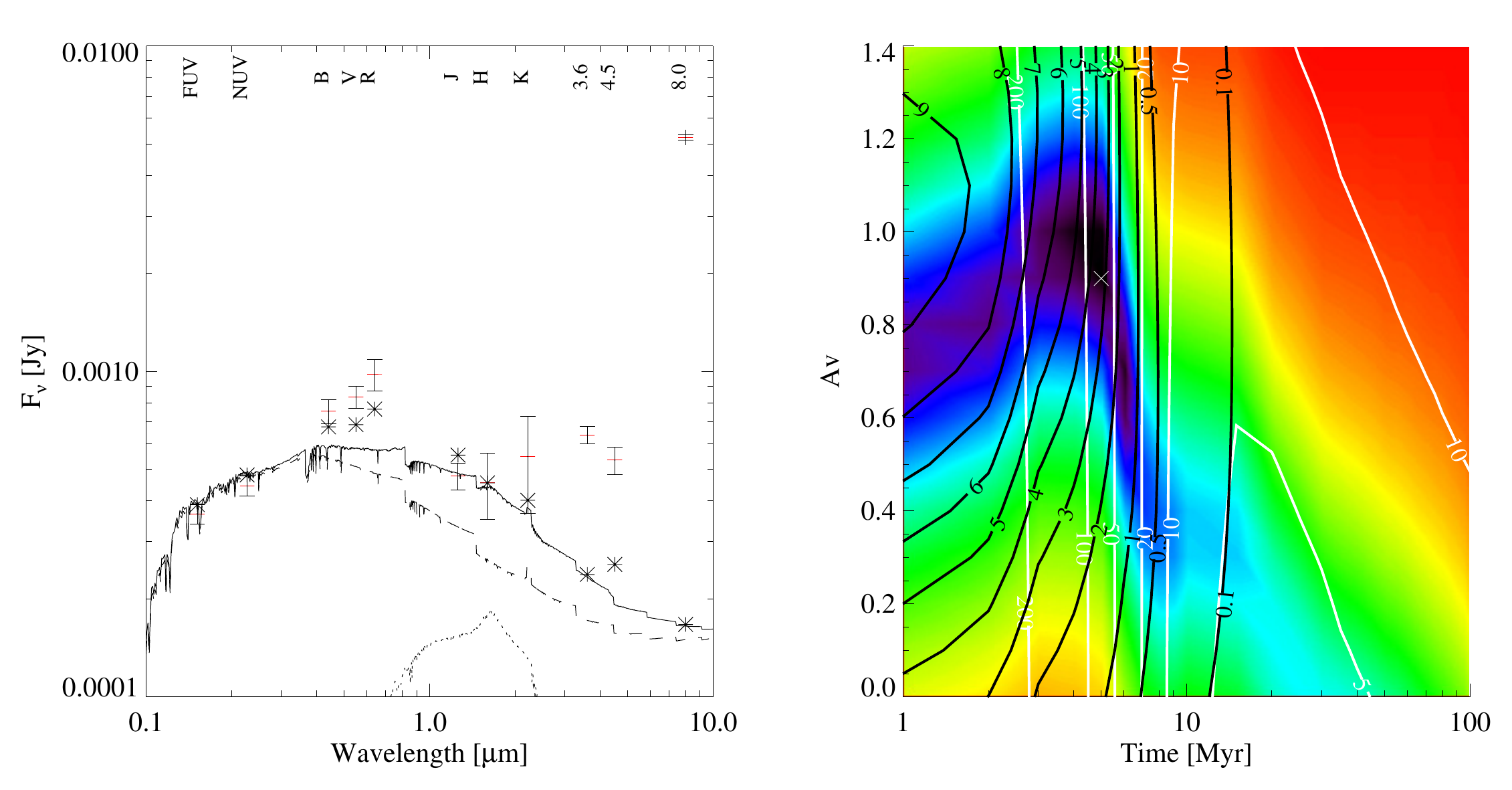}
\caption{Modeling of NGC~5291N with a mass ratio $\log r=-1$, an attenuation of 0.9 magnitude for the population formed in collision debris, no attenuation for the evolved population and a timescale $\tau=1\times10^6$ years. The fit has been performed on the FUV, NUV, B, V, R, J, H and K bands. See Figure~\ref{fig:fit-SQ-2} for additional details.\label{fig:fit-NGC5291N-mass}}
\end{figure*}

At wavelengths above 1.3 $\mu$m, the observed fluxes are 1.2--2.5 times higher than those predicted by the model. In principle, this could trace the presence of an old stellar population emitting in the near--infrared. To check this, we set $\log r=-1$. As shown in Figure~\ref{fig:fit-NGC5291N-mass}, if the fit is better for the K band, it is not improved in the other near--infrared bands and it is significantly degraded in the optical bands, the fluxes being systematically under-evaluated. If not due to stellar continuum, the excess of light in the near--infrared regime should already be attributed to dust emission, usually considered to become dominant at longer wavelength. This hypothesis will further be discussed in section~\ref{sssec:contrib-dust}.

\subsubsection{Dust}
\begin{figure*}[htbp]
\includegraphics[width=\columnwidth]{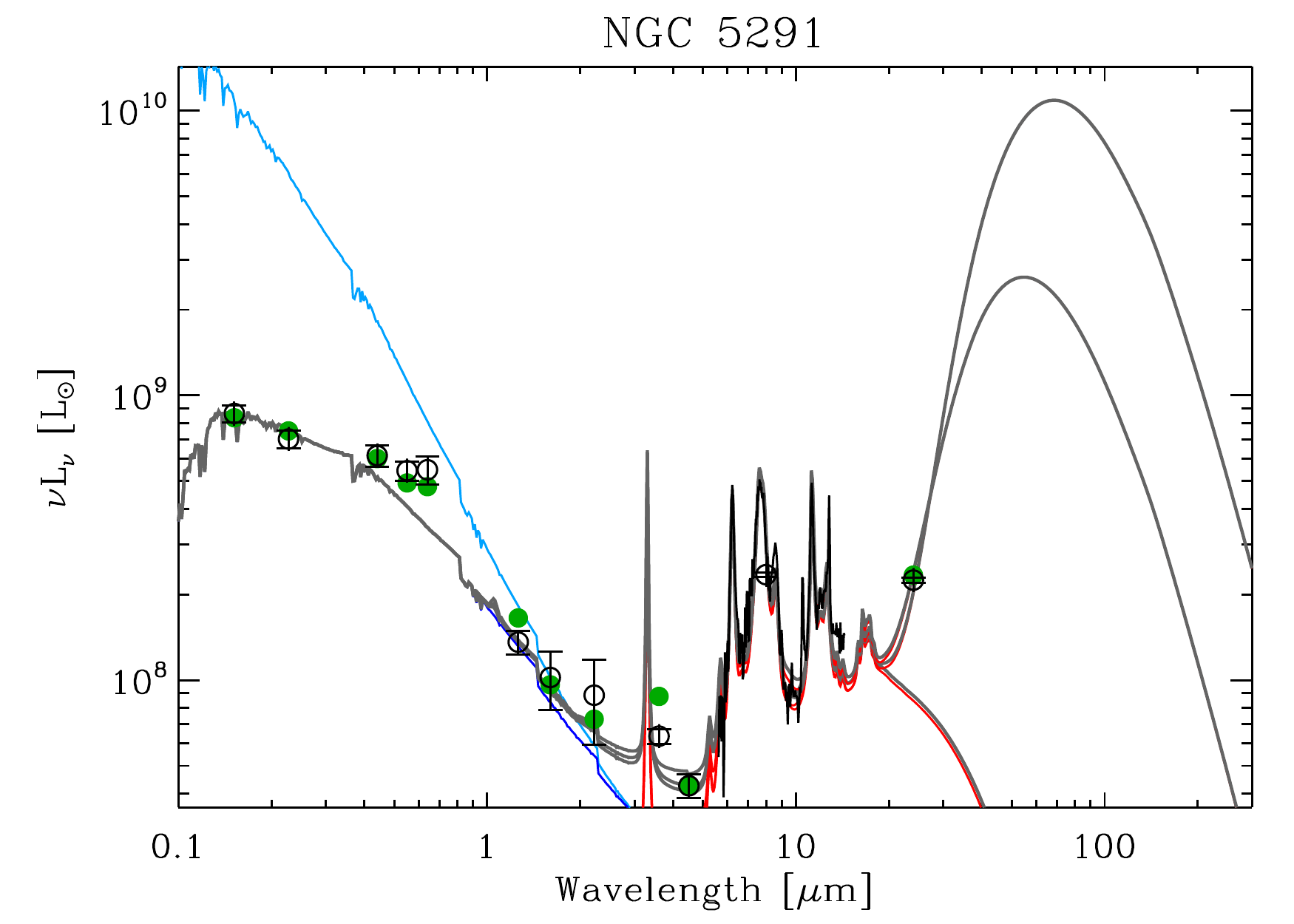}
 \includegraphics[width=\columnwidth]{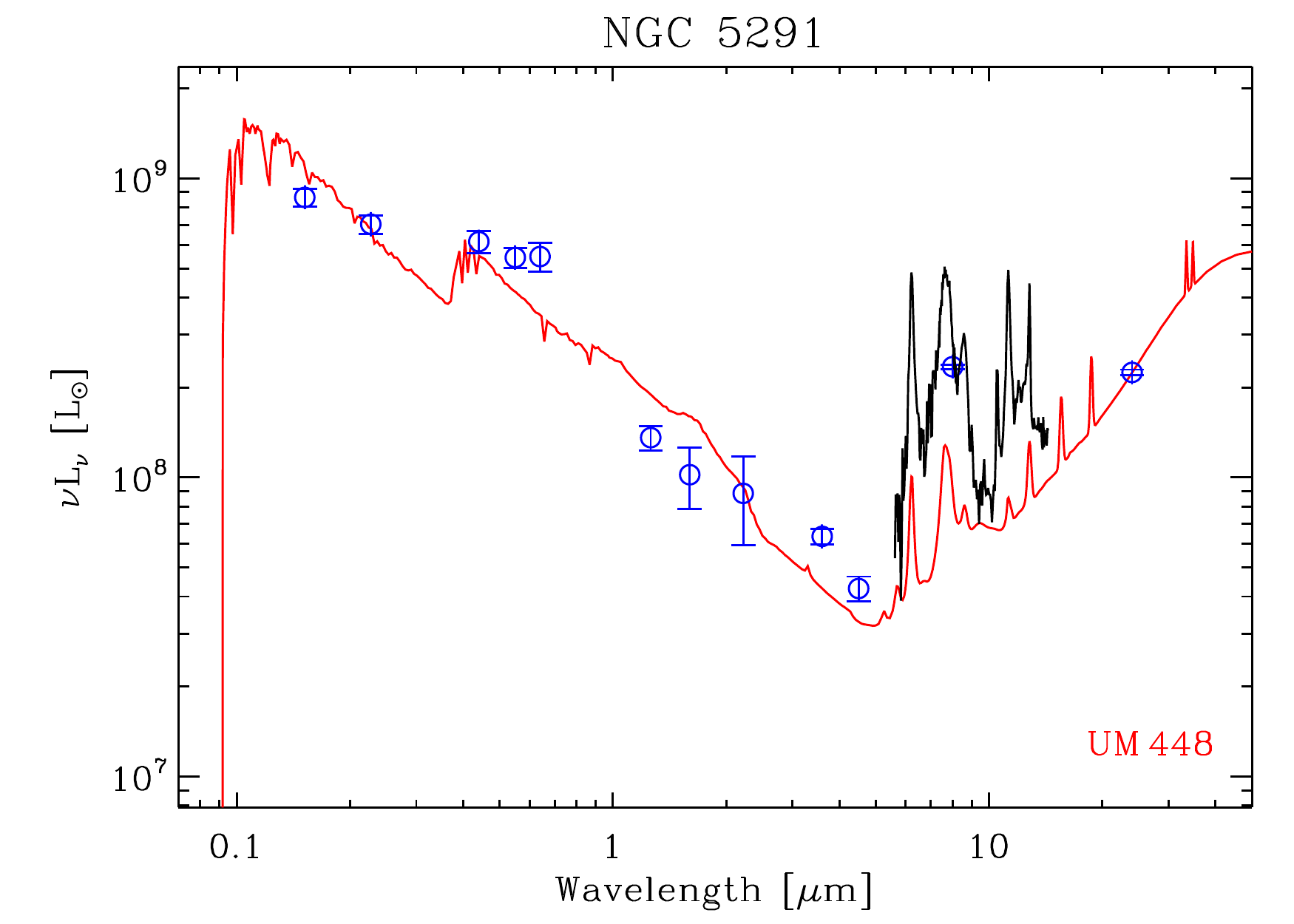}
\caption{Left: Fitted SED  of NGC~5291. The different models correspond to (1) the case where we fix $A_B=0.2$ as observed spectroscopically by \cite{duc1998a} (lowest IR emission) with $G_0=10^5$ and a starburst attenuation law, (2) we adopt the SMC attenuation curve (middle IR emission) with $G_0=445$, (3) where we use the full absorbed power constraint (highest IR emission) with $G_0=77$. The attenuation in cases (2) and (3) is derived from the mode. Right: Same as Figure \ref{fig:dust-SQ2} for NGC~5291. The black line represents the rescaled Spitzer IRS spectrum \citep{higdon2006a}.\label{fig:dust-ngc5291}}
\label{fig:sed_ngc5291}
\end{figure*}

In Figure \ref{fig:dust-ngc5291} we present the fit of the SED of NGC~5291N with its best template and a dust model.

The best galaxy template in the UV-to-optical regime is the one of UM~448 (Mrk~1304), a peculiar starburst galaxy with a relatively low metallicity (12+log(O/H)=8) but which exhibits faint PAH emission \citep{engelbracht2005a}. Like for the other active star forming regions in our sample, the emission in the mid-IR is much stronger.  NGC~5291N has more prominent PAH features. 

We fit its SED with a dust model corresponding to three different cases: 1. $A_B=0.2$ as observed by \cite{duc1998a} which corresponds to the lower infrared emission, 2. with a SMC attenuation curve which corresponds to the middle infrared emission and finally 3. using the full absorbed power constraint. The latter two cases fit similarly the data, however $A_B=0.2$ is too low to reproduce the 24~$\mu$m data.

\subsection{NGC 7252NW}

NGC~7252NW is a star-forming region located at the tip of a stellar tidal tail emanating from an old merger remnant \citep[Fig. 4 in][]{boquien2009a}.

\subsubsection{Stellar populations}

\begin{figure*}[!htbp]
\includegraphics[width=\textwidth]{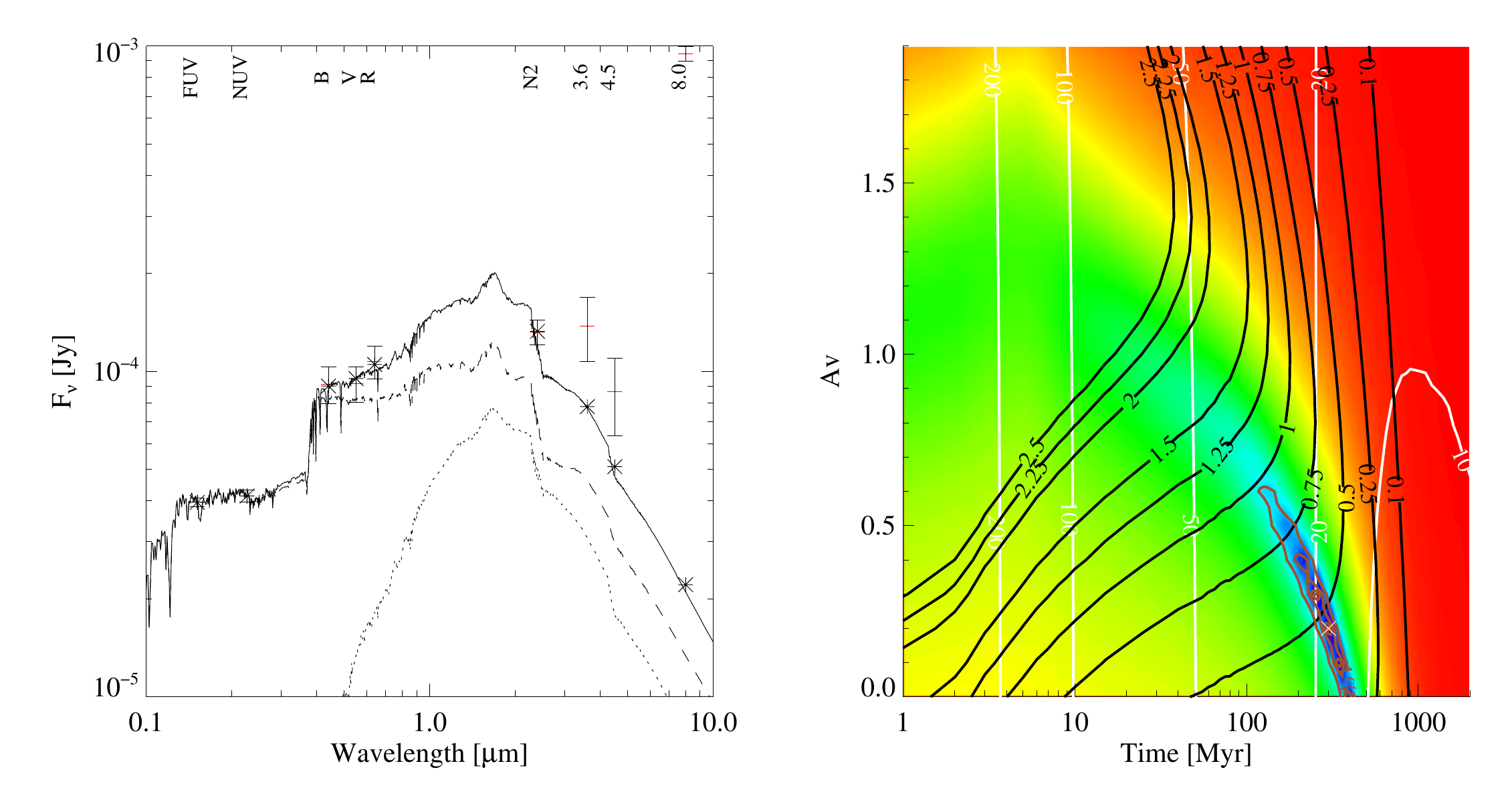}
\caption{Modeling of NGC~7252NW with a mass ratio $\log r=-0.75$, 0.9 magnitude of attenuation for the evolved population and a timescale $\tau=175\times10^6$ years. The fit has been performed on the FUV, NUV, B, V, R and N2 bands. See Figure~\ref{fig:fit-SQ-2} for additional details.\label{fig:fit-NGC7252NW}}
\end{figure*}

The best fit shown in Figure~\ref{fig:fit-NGC7252NW} reproduces the observations with a probability of 73\% with the following parameters: a rather small burst strength of $\log r=-0.75$, an age $t=300\times10^6$ years, a timescale $\tau=175\times10^6$ years an attenuation of 0.9 magnitude for the evolved population and an attenuation of 0.2 magnitude for the young population, which is significantly smaller than what was observed spectroscopically. In all cases, reasonable fits yield an attenuation lower than 1 magnitude.

\begin{figure*}[!htbp]
\includegraphics[width=\textwidth]{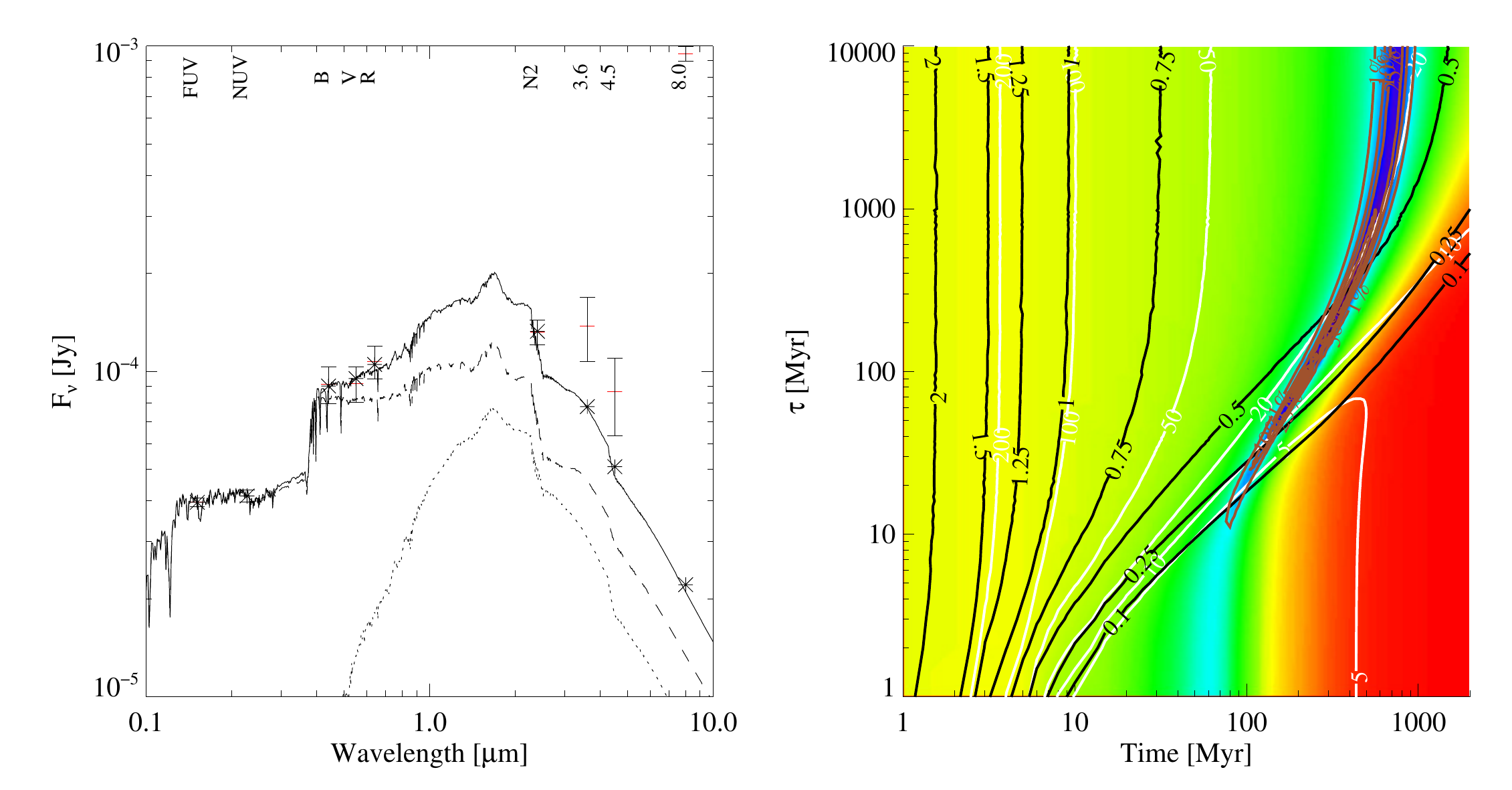}
\caption{Modeling of NGC~7252NW with a mass ratio $\log r=-0.75$, 0.9 magnitude of attenuation for the evolved population and an attenuation of 0.2 magnitude for the population formed in the collision debris. The fit has been performed on the FUV, NUV, B, V, R and N2 bands. See Figure~\ref{fig:fit-SQ-2} for additional details.\label{fig:fit-NGC7252NW-age}}
\end{figure*}

There is a wide range of possible timescales that can reproduce the observations with a probability higher than 50\% as we can see in Figure~\ref{fig:fit-NGC7252NW-age}.
A nearly instantaneous starburst leads to an age a bit younger than $60\times10^6$ years. On the contrary, for a nearly constant star formation, the best fit occurs at an age of about $t=750\times10^6$ years. However this is an upper limit at fixed $r$ and attenuation.

As the merger is old, a long timescale is not unexpected. Though, the attenuation is questionably small whereas a strong excess can be seen in the infrared.

\subsubsection{Dust}
\begin{figure*}[htbp]
\includegraphics[width=\columnwidth]{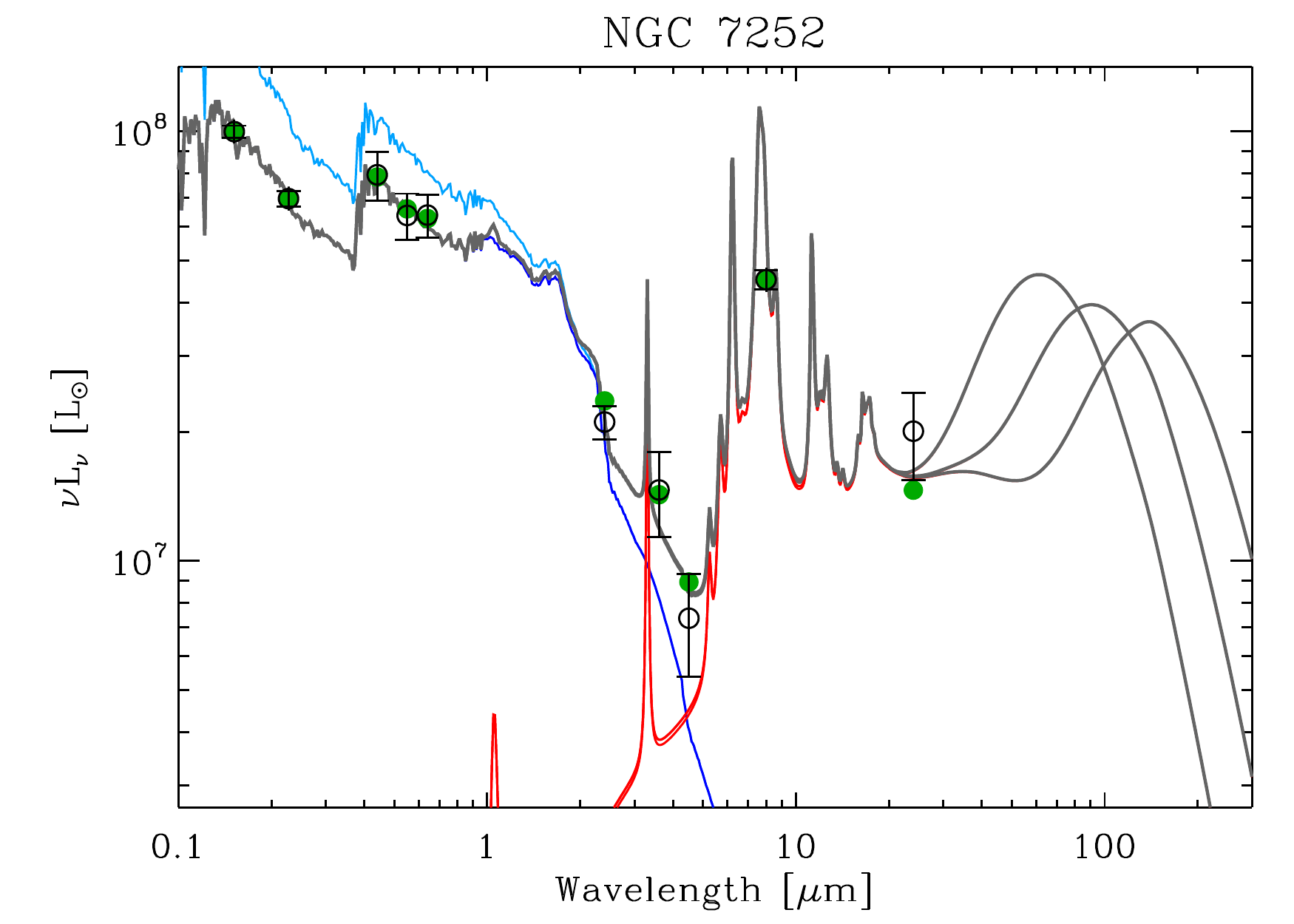}
\includegraphics[width=\columnwidth]{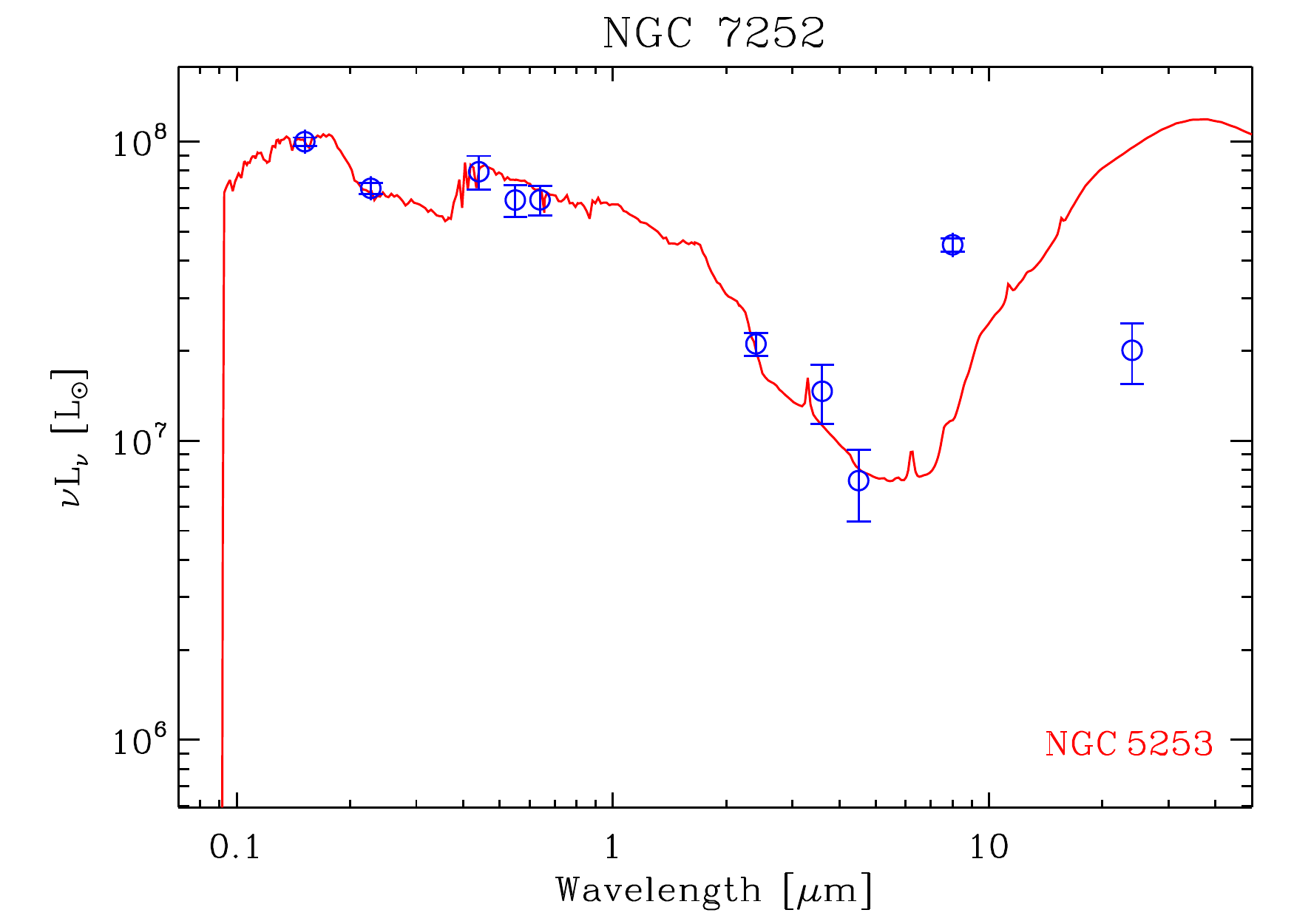}
\caption{Same as Figure \ref{fig:dust-SQ2} for NGC~7252NW. The values of $G_0$ are 1, 10 and 100.\label{fig:dust-ngc7252}} 
\end{figure*}

In Figure \ref{fig:dust-ngc7252} we present the fit the SED of NGC~7252NW with its best template and a dust model.

The best template corresponds to  NGC~5253, a nearby dwarf galaxy which has exhibited  starbursting episodes during the last few hundred million years, and where faint PAH features were detected \citep{beirao2006a}. Whereas the emission at 8 $\mu$m of NGC~7252NW is much higher, probably due to the presence of stronger PAH, its 24 $\mu$m is surprisingly lower. The dust model also shows inconsistent results. The attenuation measured from the line ratio is so low that most of the IR emission would be coming from PAH. This is highly questionable and significantly smaller than what was observed spectroscopically. This will be discussed in section \ref{sec:pahvsZ}.

\subsection{VCC~2062}
VCC~2062 has been described as an old TDG candidate, as its parent galaxy has now evolved into a relaxed early-type galaxy \citep{duc2007a}. It is linked to the latter by a star-less HI bridge.

Unfortunately this system does not have any near-infrared image available. As no diffuse stellar component that would hint about the presence of an evolved stellar population can be seen, we assume $\log r=2$. Note that a preliminary spectral energy distribution fit was published in \cite{duc2007a}.

\subsubsection{Stellar populations}

\begin{figure*}[!htbp]
\includegraphics[width=\textwidth]{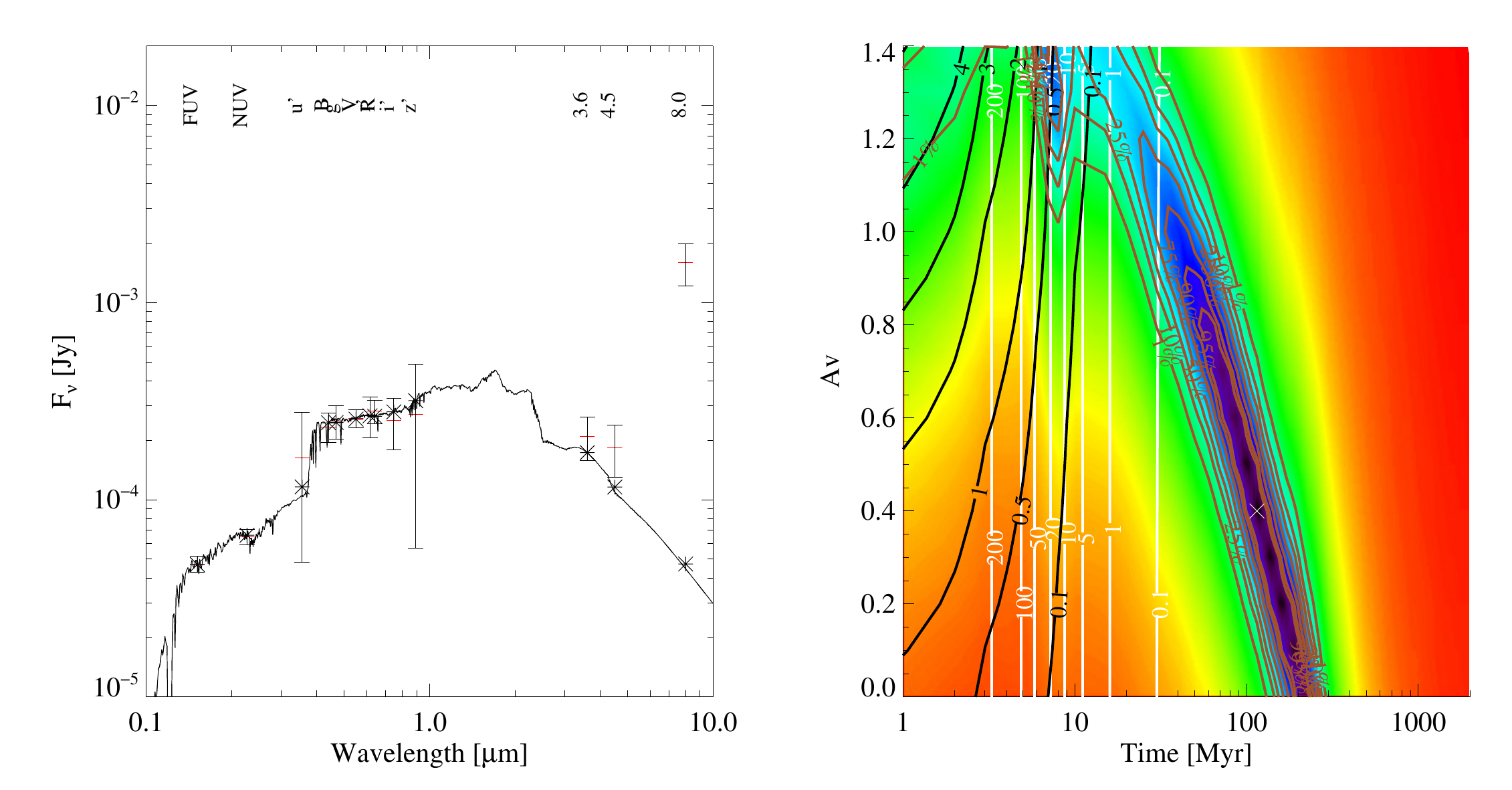}
\caption{Modeling of VCC~2062 with a mass ratio $\log r=2$ and a timescale $\tau=10^6$ years. The fit has been performed on the FUV, NUV, u', B, g', V, r', R, i' and z' bands. See Figure~\ref{fig:fit-SQ-2} for additional details.\label{fig:fit-VCC2062}}
\end{figure*}

In Figure~\ref{fig:fit-VCC2062} we present the best fit of the spectral energy distribution.
In agreement with the spectroscopic observations, the attenuation is weak, typically a few tenths of a magnitude. The best fit yields an age of $t=115\times10^6$ years and a nearly instantaneous burst: $\tau=10^6$ years.

To verify whether this result is still valid in the subregions of VCC~2062 where no more star formation can be detected, we have also performed a similar fit on the North-East part of the galaxy; there no H$\alpha$ emission can be detected. Qualitatively the results are very similar, with an age $t=270\times10^6$ years and an attenuation of 0.3 magnitude. The timescale is also very short: $\tau=2\times10^6$ years.

\begin{figure*}[!htbp]
\includegraphics[width=\textwidth]{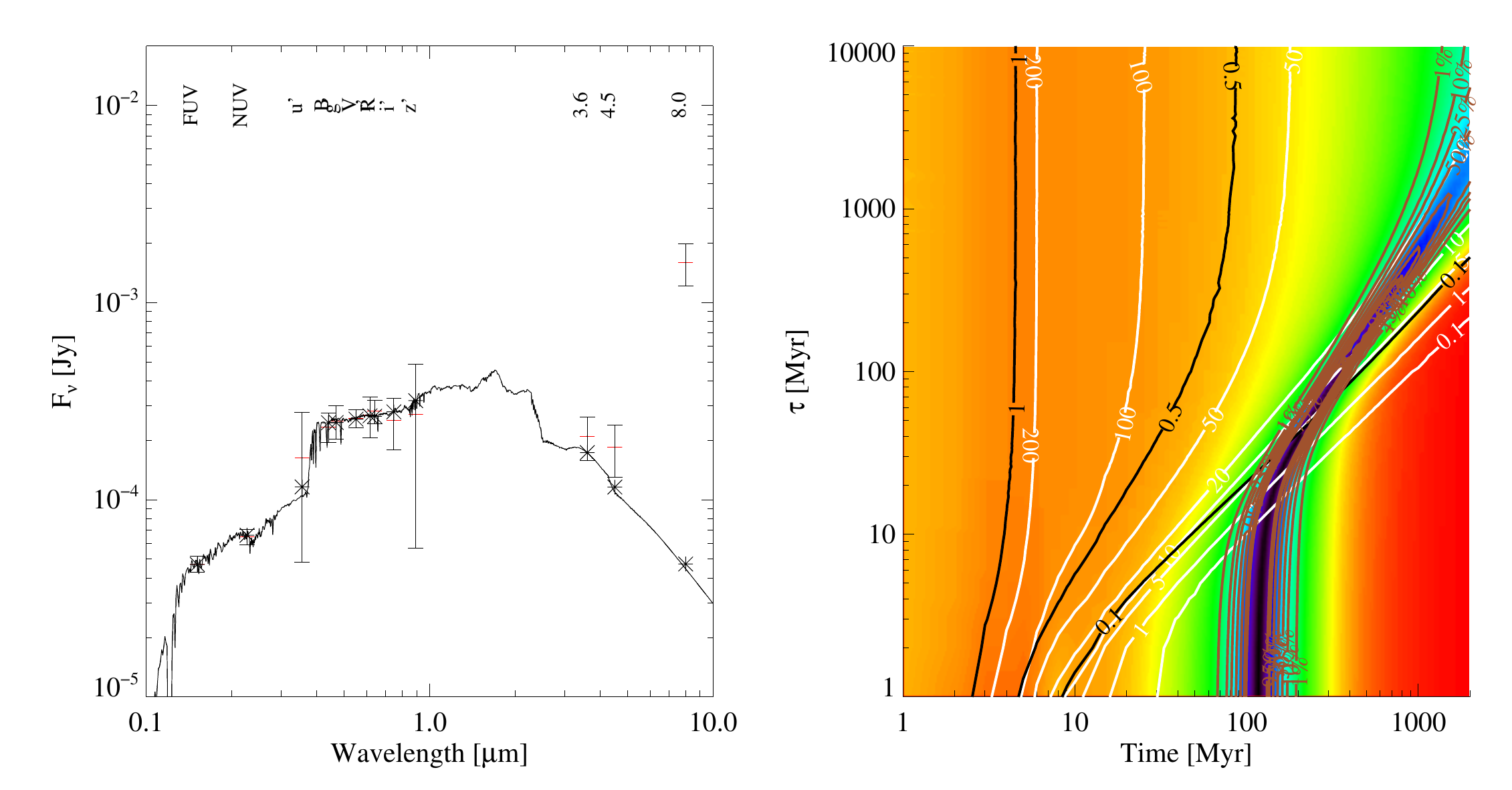}
\caption{Modeling of VCC~2062 with a mass ratio $\log r=2$ and an attenuation of 0.4 magnitude for the population formed in the collision debris. The fit has been performed on the FUV, NUV, u', B, g', V, r', R, i' and z' bands. See Figure~\ref{fig:fit-SQ-2} for additional details.\label{fig:fit-VCC2062-tau}}
\end{figure*}

We see in Figure~\ref{fig:fit-VCC2062-tau} that whatever the timescale is, the age of the burst cannot be younger than approximately $t=100\times10^6$ years.
This age changes little for a timescale shorter than a few tens of million years. However in this configuration the H$\alpha$ emission is too weak. To obtain a flux similar to the one observed, a longer timescale is necessary, about $\tau=50\times10^6$ years and an age of about $t=200\times10^6$ years. Beyond, the H$\alpha$ flux becomes significantly too high even though the fits remain proper.

Because of the uncertainties on the fluxes and the lack of near-infrared data beyond the z' band -- which flux is particularly uncertain --, it is hard to have definitive conclusions regarding this star forming region. The fits we have performed indicate that star formation started between $t=200\times10^6$ and $t=300\times10^6$ years ago. The timescale is short compared to the age, at most a few tenths of it.

\subsubsection{Dust}
\begin{figure*}[htbp]
\includegraphics[width=\columnwidth]{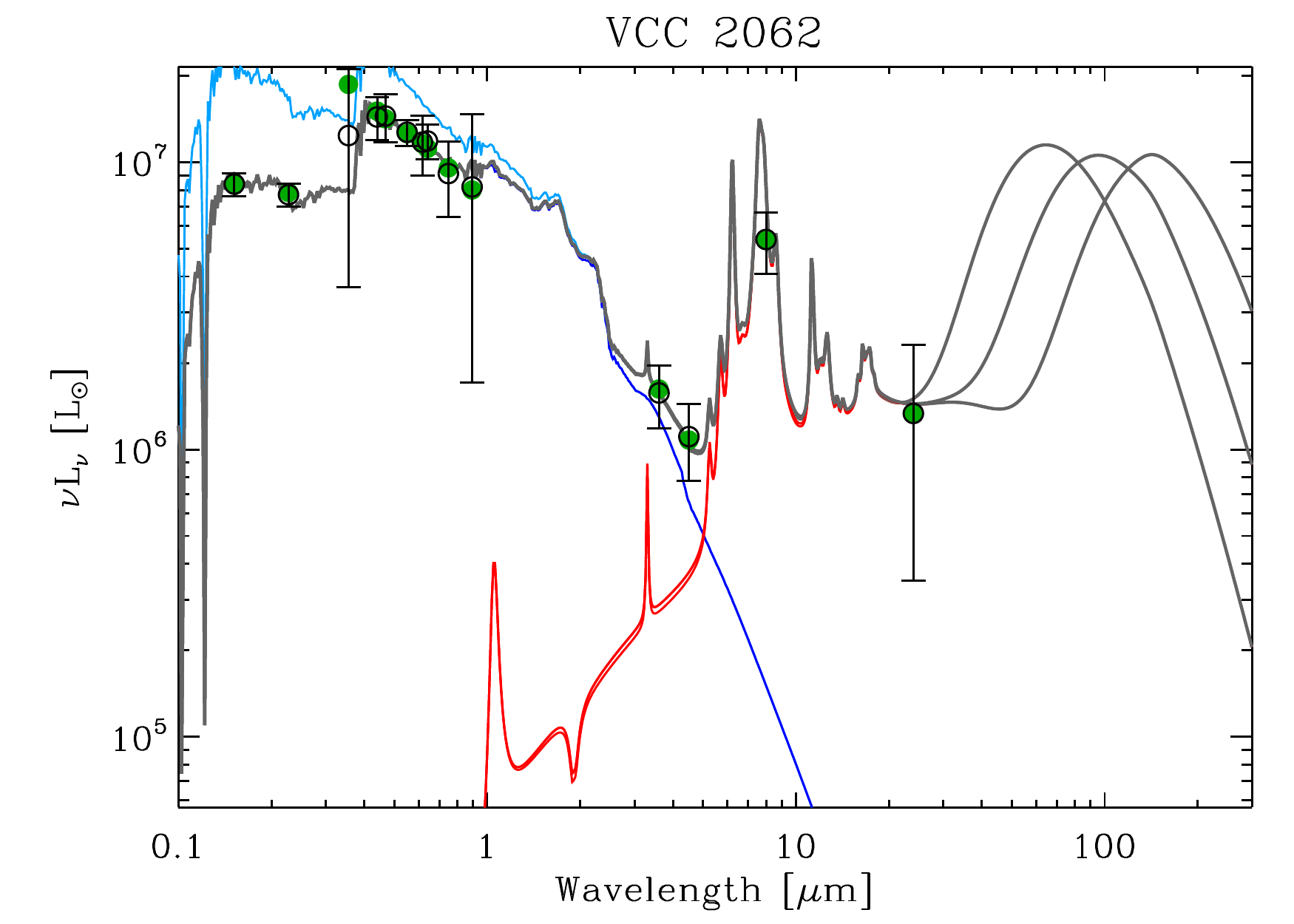}
\includegraphics[width=\columnwidth]{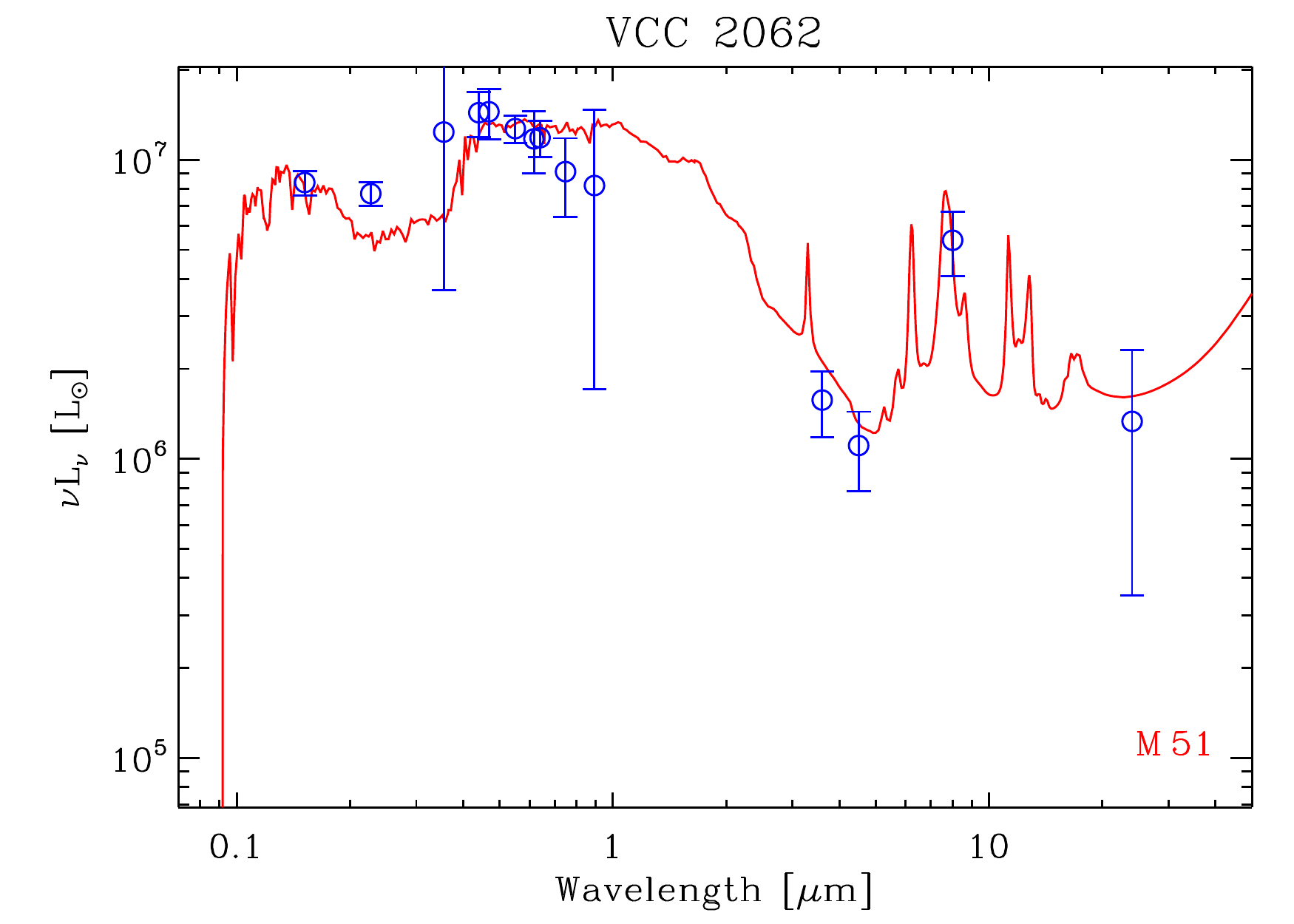}
\caption{Same as Figure \ref{fig:dust-SQ2} for VCC~2062. The values of $G_0$ are 1, 10 and 100.\label{fig:dust-vcc}} 
\end{figure*}

In Figure \ref{fig:dust-vcc} we present the fit of the SED of VCC~2062 with its best template and a dust model.

Like for SQ-2, the best galaxy template of VCC~2062 corresponds  to the spiral M51. Similarly, the fit is reasonably good from the UV to the mid-IR. 

\section{Discussion}

\subsection{Stellar populations in collision debris}

\subsubsection{Old stellar populations}

The stellar populations in tidal debris consist of an old stellar population expelled from the parent galaxies and a young one born in situ. The mass fraction of each type of stars had so far been determined qualitatively, based on the gas-richness and colors of the tidal tails: the more gas rich and the bluer they are, the higher the apparent contribution of young stars. First attempts to determine the stellar populations in a more quantitative way were made by \cite{fritze1998a} based on $B-V$, $V-R$ and $V-K$ colors. In this paper, we have used a much wider range, from the UV to the mid-IR to get more precise insights. Our study reveals a variety of properties in those objects.

Among the selected systems, four are compatible with a population constituted exclusively in the collision debris, without any underlying evolved stellar population. The conclusion is drawn from the best fit but it is also based on a series of tests as, for instance, increasing $\log r$ as was shown in section \ref{ssec:arp105}. In NGC~5291N, our result is corroborated by numerical simulations \citep{bournaud2007a} indicating that there should not be any population of galactic origin located in the collisional ring. Regarding Arp~105S, we cannot completely exclude the possibility of the presence of an evolved population even though it is unlikely. The presence of the nearby elliptical galaxy makes accurate photometric measurements difficult. In the case of SQ-5 and VCC~2062, even though fits lean towards the absence of an older stellar population, the availability of near-infrared observations is crucial to draw a more definitive answer.

Other systems such as SQ-2, Arp~245N and NGC~7252NW most likely contain an evolved stellar population, as was expected. In the case of Arp~245N, it was clearly detected and constrained thanks to near-infrared observations in the J, H and K bands.

\begin{figure}[!htbp]
 \includegraphics[width=\columnwidth]{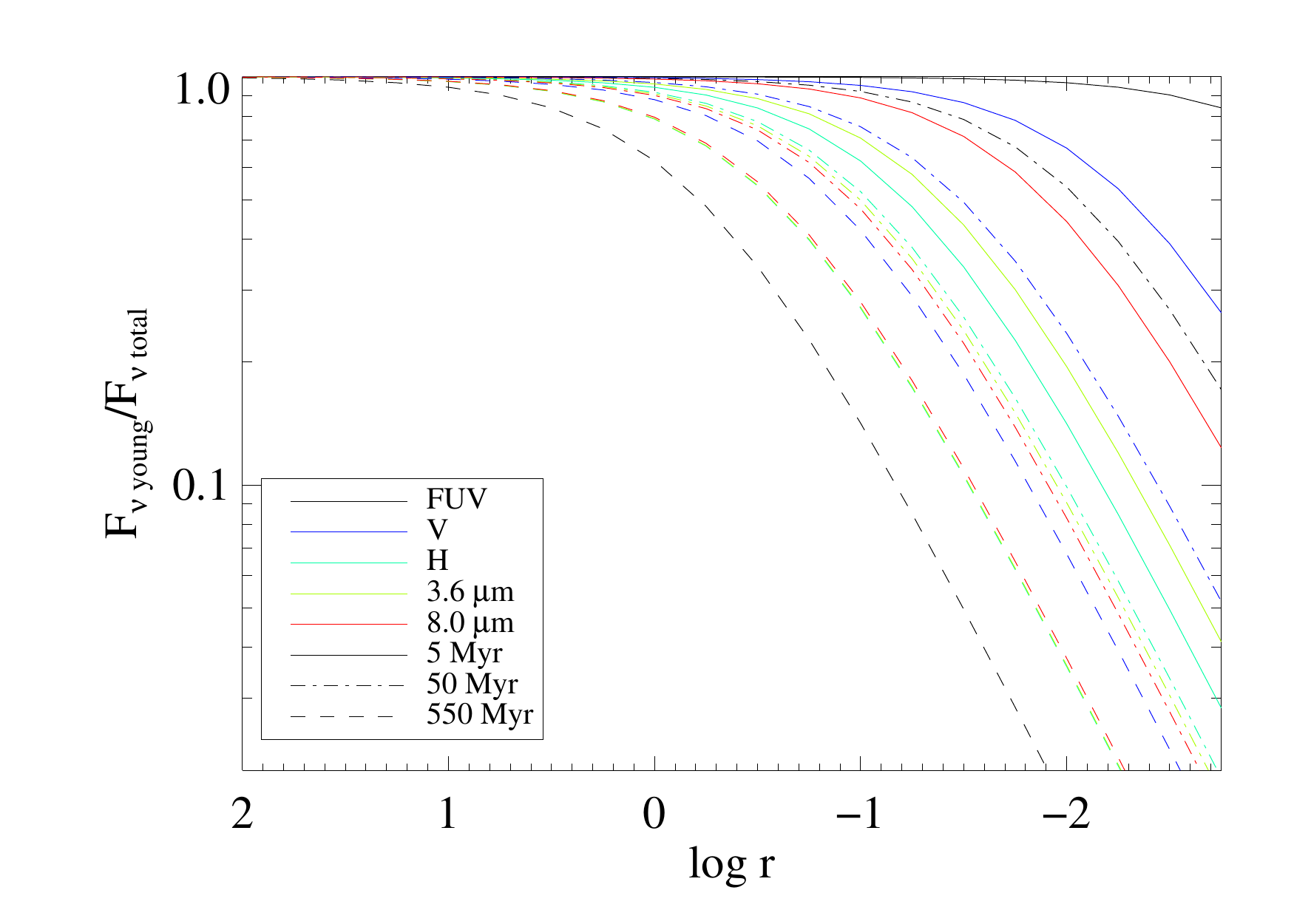}
 \caption{Evolution of the contribution of the flux in the young population relatively to the total flux in each band used during this study. The burst is nearly instantaneous et there is no attenuation. The solid line represents an age of $5\times10^6$ years, the mixed one an age of $50\times10^6$ years and the dashed one an age of $550\times10^6$ years. The bands are FUV, V, H, 3.6~$\mu$m, and 8.0~$\mu$m.\label{fig:effect-ratio-bands}}
\end{figure}

For young objects, a ratio as low as 10\% is sufficient for the young stellar population to dominate the spectral energy distribution, masking the presence of an underlying evolved stellar population as can be seen in Figure~\ref{fig:effect-ratio-bands}.

This 10\% limit depends of course strongly on the star formation history and on the quality of the available observations to disentangle the emission from both components. For a mass ratio $\log r=-1$ at $5\times10^6$ years, the young population accounts for at least 75\% of the flux, the minimum being in the H band. Yet, for the systems in which there is clearly an evolved stellar population, its emission in the near-infrared is similar to the error bars for the currently available data. Thus, for a given age, the detectability limit depends on the error bars in the near-infrared. In the present case, for a $5\times10^6$ years old burst, it is about 10\%.

\subsubsection{Recent star formation}
\label{sssec:SFR}
The recent, in-situ, star formation history in collisional debris may be be inferred comparing various tracers of star formation: ultraviolet, H$\alpha$ or mid-/far-infrared. 

Each tracer has its own characteristics. Ultraviolet is sensitive to the photospheric emission of massive stars over a timescale of a few $100\times10^6$ years, H$\alpha$ to the ionized gas surrounded star forming regions over a timescale of about $10\times10^6$ years and finally infrared is a result of the radiation of dust absorbing emission from massive young stars.

The availability of several bands permits not only to constrain the recent star formation history but also to obtain a more precise estimate of the star formation rates. However they are still sensitive to some degeneracies, depending on the observations available, and thus the fits must be cautiously interpreted. We compare the SFR obtained from the SED fitting, presented in Table~\ref{tab:fit-parameters} to the ones published in \cite{boquien2009a} which use the conversion factors published by \cite{kennicutt1998a} for the UV and the H$\alpha$ emission and by \cite{wu2005a} for the 8~$\mu$m. The SFR given by the models are the ``true'' instantaneous SFR. The ones published in \cite{boquien2009a} assumed a given star formation history.

We notice that the model strongly underestimates the SFR from the observations for VCC~2062 if we assume a single recent burst. This is most likely due to several factors: the lack of observations in near-infrared beyond the z' band and a star formation history that is not properly reproduced by our set of models. An older age of about 300~Myr with a timescale around 100~Myr, which is not ruled out by the model is probably more realistic as it would take into account more accurately a long term low level star formation. For the other models, except for NGC~5291N, SFR(FUV) systematically underestimates SFR(model). In the case of SQ-5, SFR(model)/SFR(FUV) even reaches $\sim32$. This is chiefly an attenuation effect: spectroscopic observations of NGC~5291N show that the attenuation is very small \citep{duc1998a} and SQ-5 is particularly bright at 8~$\mu$m compared to FUV and has a relatively high attenuation. Regarding ionized gas, SFR(model)/SFR(H$\alpha$) is comprised between 0.73 for NGC~7252 and 2.8 for SQ-5. No trend can be seen with the star formation history or the attenuation hinting at a complex combination of several factors. Finally, SFR(model) and SFR(8.0) -- the star formation rate obtained from the PAH emission in the Spitzer 8.0~$\mu$m band -- provide similar results with two exceptions: SQ-5 reaches a ratio of $\sim4.5$ and Arp~245N presents a strong mid-infrared emission with a ratio of 0.28.

Even though the standard SFR estimators often yield results that have the correct order of magnitude, an error of several times is common with extreme cases of up to several tens. This shows that as SED fits allow to model the star formation history and the attenuation they permit to obtain more accurate results. The SFR from the H$\alpha$ provides relatively good estimates of the real star formation rate, with a maximum excursion of 2.44. This is not surprising as the timescale of H$\alpha$ emission is small and therefore more closely traces recent star formation.

\tabletypesize{\tiny}
\begin{deluxetable*}{cccccccccc}
\tablecolumns{16}
\tablewidth{0pc}
\tablecaption{Best fit parameters}
\tablehead{
\colhead{System}&\colhead{$t$}&\colhead{$\tau$}&\colhead{$A_V$}&\colhead{$\log r$}&\colhead{Total stellar mass}&\colhead{SFR}&\colhead{SFR/SFR(UV)}&\colhead{SFR/SFR(H$\alpha$)}&\colhead{SFR/SFR(8.0)}\\
\colhead{}&\colhead{years}&\colhead{years}&\colhead{magnitudes}&\colhead{dex}&\colhead{M$_\sun$}&\colhead{M$_\sun$ yr $^{-1}$}&}\\
\startdata
Stephan's Quintet 2&$85\times10^6$&$37\times10^6$&$1.1$&$-1$*&$178\times10^6$&0.08&2.32&2.10&1.09\\
Stephan's Quintet 5&$2\times10^6$&$10^6$ -- $10\times10^9$&$1.2$&$2$*&$674\times10^3$--$696\times10^3$&0.21&31.62&2.76&4.45\\
Arp~105S&$4\times10^6$&$10^6$&$1.6$&$2$&$66\times10^6$&0.28&1.77&0.76&0.99\\
Arp~245N&$115\times10^6$&$10^6$&$1.5$&$-0.75$&$1.2\times10^9$&0.02&1.51&0.90&0.28\\
NGC~5291N&$5\times10^6$&$10^6$ -- $10\times10^9$&$1.1$&$2$&$29\times10^6$--$33\times10^6$&0.14&0.61&0.73&0.85\\
NGC~7252NW&$300\times10^6$&$175\times10^6$&$0.2$&$-0.75$&$94\times10^6$&0.03&1.07&0.75&0.89\\
VCC~2062&$115\times10^6$&$10^6$&$0.4$&$2$*&$47\times10^6$&$1.18\times10^{-6}$&$5.17\times10^{-4}$&$1.02\times10^{-3}$&$3.04\times10^{-4}$
\enddata
\tablecomments{The ``*'' symbol indicates that the parameter was fixed prior performing the fit.\label{tab:fit-parameters}}
\end{deluxetable*}

\subsection{Dust in collision debris}

As star--formation starts in dust cocoons, the dust component should be an important ingredient of collisional debris and indeed it has been observed in this environment either in absorption, as dust lanes, that can be seen for instance in the region B of Stephan's Quintet \citep{lisenfeld2004a}, or directly in emission \citep{higdon2006a,boquien2009a}. Various types of dust may be probed by our data, as discussed below.

\subsubsection{A qualitative comparison with UV-to-IR galaxy SED templates}
\label{sec:dust}

In section \ref{sec:results}, we have shown the comparison of the observed SED of each star forming region with the SED of nearby galaxies modeled by \citet{galliano2008a}. Their 35 templates were scaled to match the UV-to-near-IR (up to 4.5~$\mu$m) observed SED of the galaxies presented in this study. We kept the template minimizing the $\chi^2$.

For Arp~105S, Stephan's Quintet 5, NGC~7252NW and NGC~5291N, the templates that match best the stellar spectrum are templates of low-metallicity, galaxies. Indeed, the stellar SED of these objects is very blue. However, when comparing the fitted template to the mid-IR observations, the observations present an excess emission. This indicates that the PAH abundance is higher than in the templates. High PAH abundance is the sign of a metal rich ISM. Therefore, those 4 intergalactic star forming regions have a young stellar population typical of a blue compact galaxy, but have the mid-IR emission of a solar metallicity galaxy.

For the rest of the sample, Arp~245S, Stephan's Quintet 2, and VCC~2062, are relatively well matched by SED templates of spiral galaxies, because their stellar spectrum is redder.

\subsubsection{The PAH-to-dust mass ratio and the metallicity}
\label{sec:pahvsZ}

The PAH-to-dust mass ratio is a function of ISM metallicity. \citet{madden2006a} and \citet{wu2006a} have shown that the PAH feature strength to continuum intensity ratio is correlated both with the metallicity of the ISM and the [Ne$\,${\sc iii}]/[Ne$\,${\sc ii}] line ratio. The latter is a tracer of very young stellar populations, and therefore an indicator of the hardness of radiation field. \citet{madden2006a} proposed that the paucity of PAH in low-metallicity environments was a consequence of their large-scale destruction by the hard penetrating radiation. Indeed, those environments contain more ionizing radiation, and their ISM has a lower opacity, due to the lower dust abundance.

\citet{galliano2008a} showed that although PAH destruction is important, the trend is governed by the delayed injection of PAH by their progenitors, the AGB (asymptotic giant branch) stars. AGB stars are believed to be the main source of PAH. AGB stars begin to inject their elements and dust after their lifetime of $\simeq400$~Myr, while massive stars inject their elements and dust (non-PAH carbon and silicate dust) into the ISM after only a few $10^6$ years. The metallicity being a function of the age of the galaxy, the delayed injection of PAH by AGB stars explains the deficit of PAH in low-metallicity environments.

Although the origin of this trend is still controversial, the trend in itself is consensual. We can therefore assume that the PAH-to-dust mass ratio scales with the metallicity of each of our intergalactic star forming regions.

\begin{figure}[htbp]
\includegraphics[width=\columnwidth]{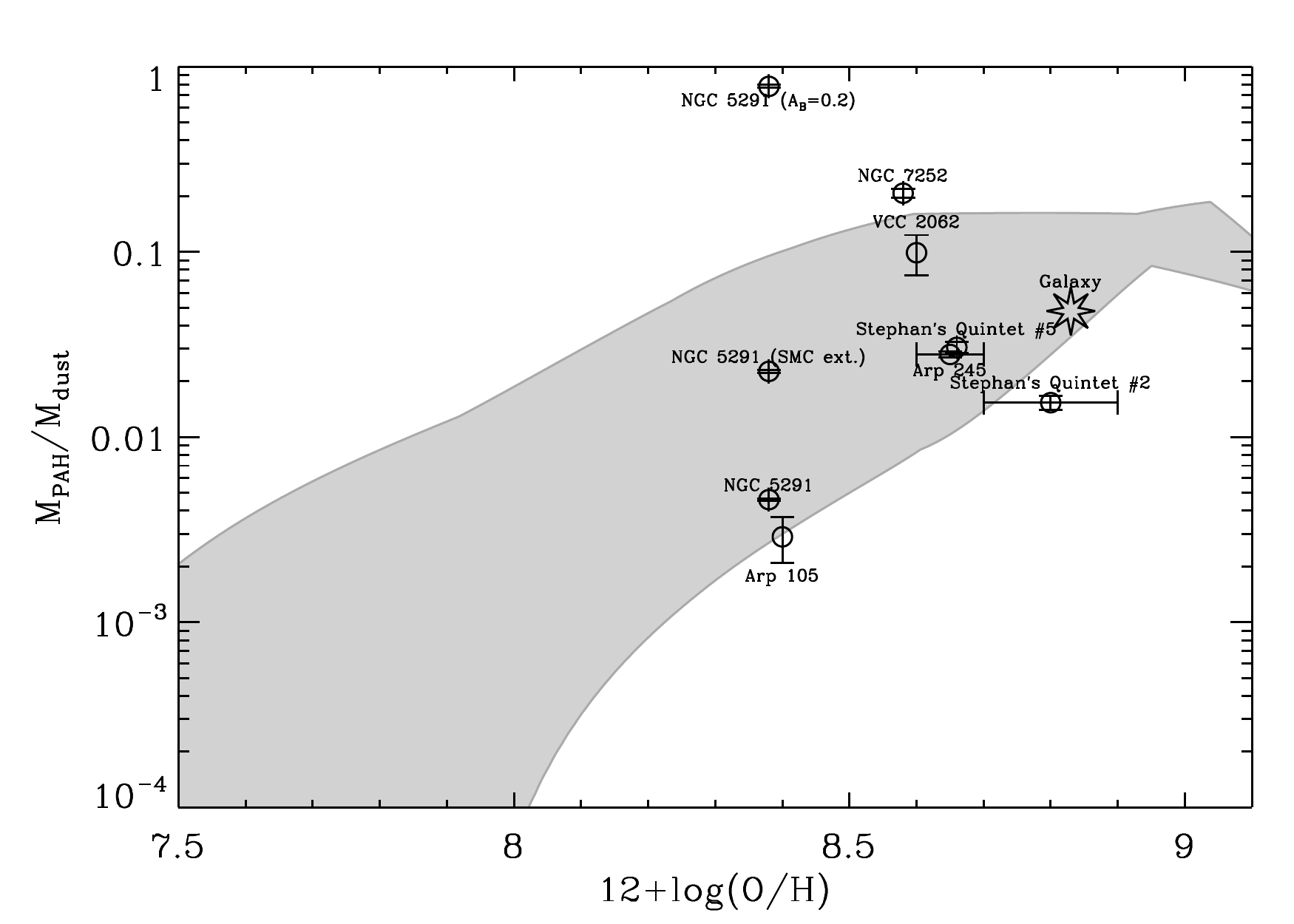}
\caption{Comparison the derived PAH-to-dust mass ratio to their theoretical value. The points with error bars are the PAH-to-dust mass ratios derived from the SED fit, as a function metallicity. The gray curve is the theoretical value of this ratio, from chemical evolution considerations \citep{galliano2008a}.}
\label{fig:pahvsZ}
\end{figure}

As a test of the attenuation obtained with the photospheric+nebular model, Figure \ref{fig:pahvsZ} compares the values of the PAH-to-total dust mass (comprising the PAH, the Very Small Grains and the Big Grains) ratio $f_{PAH}$ to the range of values consistent with the observed trend with metallicity \citep{galliano2008a}. We emphasize that this value of $f_{PAH}$ relies on the accuracy of the attenuation estimate. Here, we do not consider this determination of $f_{PAH}$ as an actual measure of the PAH-to-dust mass ratio, but we rather compare this derived $f_{PAH}$ to its expected value, taking the trend of PAH abundance with metallicity for granted. From this comparison, there are three configurations. First, if these two values are in agreement, it provides an additional indication that the attenuation estimate, as well as the fitted stellar populations, might be accurate. Second, if the derived $f_{PAH}$ is much higher than its expected value, it is the indication that the attenuation estimate implies a gross underestimate of the IR luminosity. It therefore suggests that regions contributing to the IR luminosity were not probed in attenuation by the photospheric+nebular fit. These regions are likely embedded star forming regions, contributing to the IR power, but optically thick to visible recombination lines. Finally, if the derived value of $f_{PAH}$ was lower than its expected value, it would mean that we overestimated the attenuation. This situation would be more difficult to justify from a physical point of view. However, we never encounter this situation in our sample.

Most objects are within the expected range, except NGC~5291N and NGC~7252NW, in the case where we adopt $A_B=0.2$ as was observed spectroscopically by \cite{duc1998a}. The fact that these two objects are outliers is an indication that the attenuation estimate does not account for enough IR emission. It is obvious from their SED themselves (Figures~\ref{fig:dust-ngc5291} and \ref{fig:dust-ngc7252}, left panel), where we see that for those two cases, the IR emission comes essentially from the PAH. This discrepancy allows us to estimate that there is a missing IR luminosity, likely coming from embedded star forming regions. For NGC~5291N and NGC~7252NW, this missing IR luminosity is about an order of magnitude higher than the power derived from the attenuation estimate. Table \ref{tab:pahvsZ} summarizes those parameters.

\tabletypesize{\normalsize}
\begin{deluxetable*}{ccccc}
\tablecolumns{5}
\tablewidth{0pc}
\tablecaption{Parameters relative to the mid-IR SED fit.}
\tablehead{
\colhead{Object}&\colhead{$M_{PAH}/M_{dust}$}&\colhead{Absorbed power $[L_\sun]$}&\colhead{Missing $L_{IR}\;[L_\sun]$}}\\
\startdata
Arp$\,$105S & $0.003 \pm 0.001$ & $3.1\times 10^{10}$ & \ldots \\
Arp$\,$245N & $0.028 \pm 0.001$ & $10^{9}$ & \ldots \\
Stephan's Quintet \#2 & $0.015 \pm 0.001$ & $10^{9}$ & \ldots \\
Stephan's Quintet \#5 & $0.030 \pm 0.002$ & $5.1\times10^{8}$ & \ldots \\
NGC$\,$7252NW & $0.21 \pm 0.01$ & $10^{8}$ & $\simeq$~$10^{8}$~--~$2\times10^{9}$ \\
VCC$\,$2062 & $0.10 \pm 0.02$ & $10^{7}$ & \ldots \\
NGC$\,$5291N $(A_B=0.2)$ & $0.78 \pm 0.01$ & $2.9\times10^{8}$ & $\simeq$~$2\times10^{9}$~--~$8\times10^{10}$ \\
NGC$\,$5291N & $0.0046 \pm 0.0001$ & $10^{10}$ & \ldots \\
NGC$\,$5291N (SMC ext.) & $0.023 \pm 0.0004$ & $3.0\times10^{9}$ & \ldots \\
\hline
Galaxy & 0.046 & \ldots & \ldots
\enddata
\label{tab:pahvsZ}
\end{deluxetable*}

We also note that there is an anticorrelation between M(PAH)/M(dust) and Av. A possibility would be that it is due to the color of the ISRF. However, it is lower than the variations we are probing. For example, Fig. 3 of \cite{galliano2008a} demonstrates the effect of the color of the ISRF on the PAH spectrum, for two extreme cases. The variation is at most a factor of 3, and it is certainly lower in this sample: we are probing variations of 2 order of magnitude in M(PAH)/M(dust). The reason of this anticorrelation most likely reflects the systematic effect of the determination of Av. If we consider the V band flux over 8~$\mu$m in our sample, it does not vary drastically: $F_\nu(V)/F_\nu(8)\simeq0.1$, SQ5 being lower (0.02) and NGC 245 higher (0.25). At first order, the value of Av will determine the fraction of absorbed luminosity in the UV-visible, relative to the V band. Since, the V band to 8~$\mu$m flux does not vary a lot, Av will determine the absorbed luminosity relative to the 8~$\mu$m band. Finally, the value of M(PAH)/M(dust) will depend directly on the 8~$\mu$m to absorbed luminosity ratio. Therefore, when Av is low, 8~$\mu$m/absorbed luminosity is going to be high, and therefore, M(PAH)/M(dust) is going to be high. Fig.~\ref{fig:pahvsZ} is a way to help discriminate the values of Av which are clearly too low compared to the known 8~$\mu$m emission. A very low value of Av (e.g. NGC 5291 Av=0.2) would require all the observed luminosity to be put in the PAHs. There would be no room for far-IR dust emission (Fig.~\ref{fig:fit-NGC5291N-mass}), which is difficult to understand theoretically, and has never been observed at the scale of a galaxy.

\subsubsection{Dust contribution to the 3.6~$\mu$m and 4.5~$\mu$m bands}
\label{sssec:contrib-dust}

The modeling of the fluxes at 3.6~$\mu$m and 4.5~$\mu$m by the photospheric and nebular models is not always very satisfying, in particular for the regions for which the luminosity is dominated by the population formed in the collision debris.
\begin{figure*}[!htbp]
\includegraphics[width=\columnwidth]{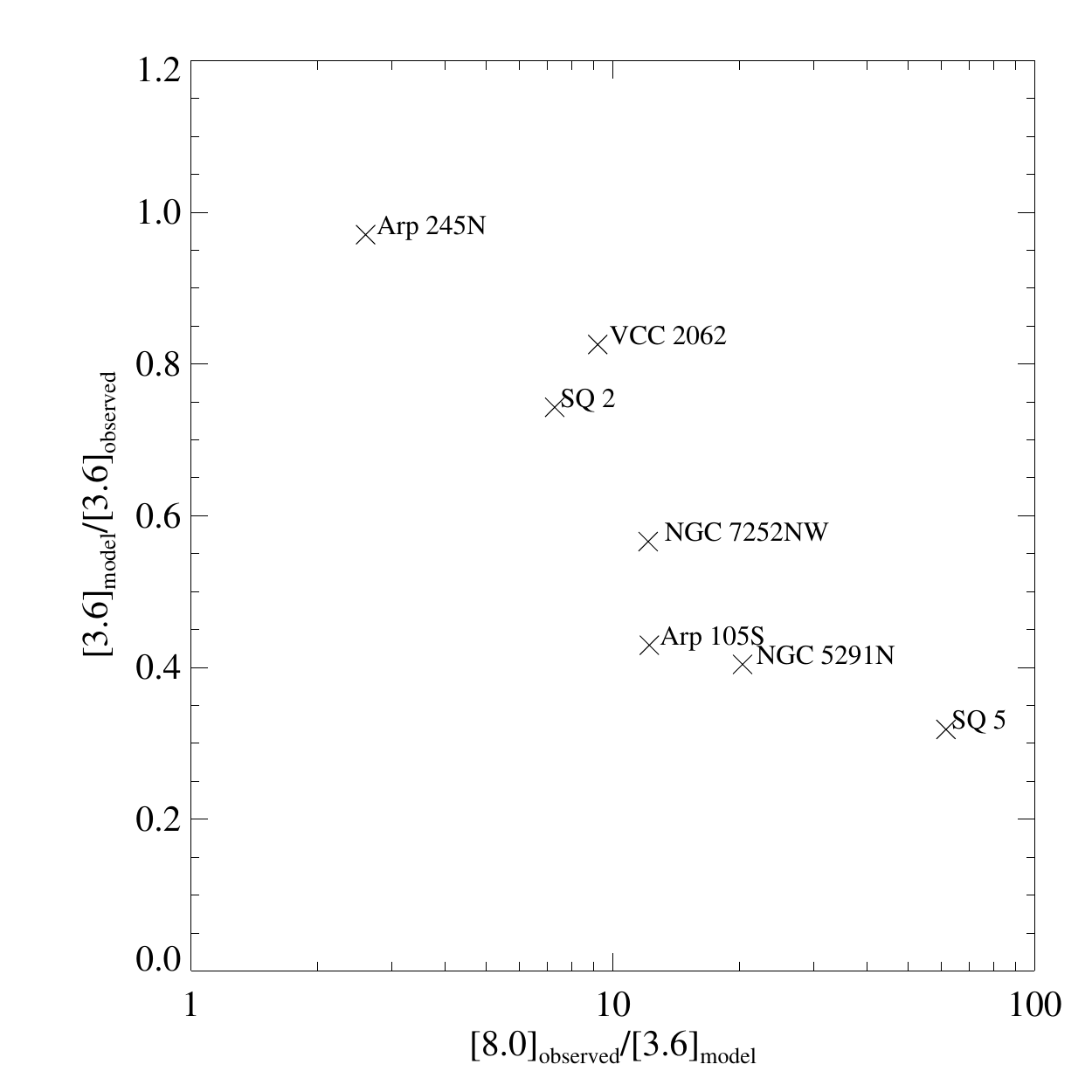}
\includegraphics[width=\columnwidth]{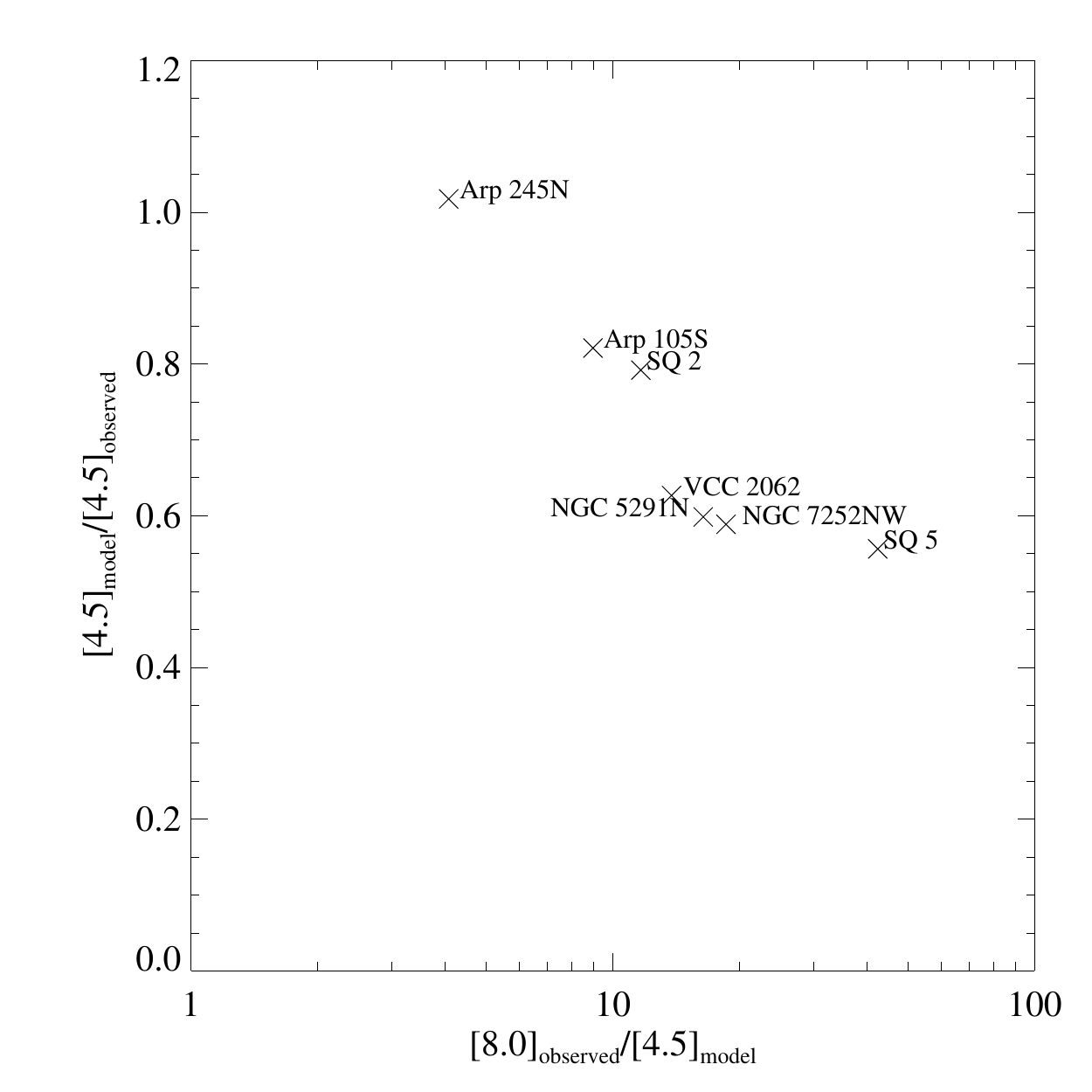}
\caption{Left: Ratio of 3.6~$\mu$m model flux to the 3.6~$\mu$m observed flux versus the specific star formation rate. Right: same as left but for the 4.5~$\mu$m band. The fluxes are obtained from photospheric and nebular only models, not from the dust models.\label{fig:fit-quality}}
\end{figure*}
To exhibit the physical cause, we have plotted the ratio of the modeled to the observed bands versus the observed 8~$\mu$m band emission (which traces star formation) normalized to the emission of the model at 3.6~$\mu$m, which can be seen as a specific star formation rate.
In Figure~\ref{fig:fit-quality} we see that as the specific star formation rate increases, the fit of the 3.6~$\mu$m and 4.5~$\mu$m bands degrades progressively. The Pearson correlation coefficient is $r=-0.71$ (resp. $r=-0.72$) for the 3.6~$\mu$m (resp. 4.5~$\mu$m) band.

\cite{smith2009a} have made an extensive study on the origin of the non-stellar 4.5~$\mu$m excess in dwarf galaxies investigating several possibilities. They have shown this excess is likely a combination of nebular emission (Br-$\alpha$ line and continuum) and high optical depth -- which would redden the 3.6~$\mu$m to 4.5~$\mu$m color --, making the presence of dust emission unnecessary. However we cannot rule out that the contamination of the 3.6~$\mu$m and 4.5~$\mu$m bands by dust emission is a possibility in star forming regions in collision debris. A high optical depth may be a factor in NGC~5291N and NGC~7252NW as we have shown that there are hints of embedded star forming regions. Our UV to near-infrared model already takes into account the nebular emission, both the Br-$\alpha$ line and the continuum, so the detected excess is unlikely to be of nebular origin in its entirety. This leads us to interpret this excess as the effect of the emission of the dust (the continuum as well as the emission lines) excited and heated by massive stars. This emission is probably coming from very small grains fluctuating in temperature, rather than hot grains, because the equilibrium temperature of the latter ($\simeq$700K) would correspond to grains too close to stellar sources ($\simeq$1 a.u. of an O4 star). When the specific star formation rate is high, there are proportionally many more young stars that are able to make the small grains fluctuate to higher temperatures than the rest of the stellar population. As a consequence, the dust emission at short wavelength becomes non-negligible compared to the stellar emission. As a consequence, the 3.6~$\mu$m and 4.5~$\mu$m bands cannot be good tracers of the mass of the evolved stellar population, especially in regions that have a high specific star formation rate. In addition, the objects that have the highest excess at 3.6~$\mu$m and 4.5~$\mu$m are also those with a negligible evolved stellar population that can contribute to those bands, making it easier for the dust to dominate. An uncertainty may be introduced by the $r$ parameter. It should have little influence as the J, H and K near-infrared bands strongly constrain the modeled SED in longward bands in wich the stellar emission is dominated by the same populations. However, when no near-infrared band is available shortward of 3.6~$\mu$m, the $r$ parameter is much harder to constrain and the uncertainty in the 3.6~$\mu$m and 4.5~$\mu$m bands is much higher.

\section{Summary and conclusions}

In this study we have compiled and modeled the spectral energy distribution of a sample of 7 star forming regions in collision debris. We have carried out this study in three steps. First, we have adjusted the SED from the ultraviolet to the near-infrared (if available, otherwise ultraviolet to optical) with a photospheric+nebular model, yielding information on the star formation rate, attenuation, and star formation history. In a second step, we have fitted their UV to near-IR SED with templates of well known galaxies and then compared the mid-infrared part of their SED to infer information on the dust content. Finally, a dust model was made to obtain estimates of the dust-to-PAH mass ratios. We have obtained the following main results:

\begin{enumerate}
 \item We confirm that several star forming regions seem compatible with a complete absence of any evolved stellar population: Stephan's Quintet 5, Arp~105S and NGC~5291N. This makes them ideal to study star formation isolated from the influence of any evolved stellar population.
 \item The UV-to-optical SED of these star forming regions exhibiting high burst strengths are best fit with templates of starbursting dwarf galaxies, while their near to mid-IR SED resemble more that of dusty, metal-rich, star-forming regions in spiral disks. The collision debris dominated by an old stellar component, like Stephan's Quintet 2, show a good fit with spiral disks from the UV to the mid-infrared.
  \item The 3.6~$\mu$m and 4.5~$\mu$m bands have a luminosity that cannot be explained by the stellar luminosity alone. This excess is correlated with the star formation intensity. The contamination most likely originates from very small grains. As such, those bands should not be used to trace the stellar mass in collision debris unless the star formation intensity is weak.
 \item The 8~$\mu$m emission provides an additional consistency check, showing that most of our attenuation estimates and subsequent stellar population synthesis are reliable. We find discrepancies in two cases: NGC~5291N and NGC~7252NW. The nature of these discrepancies indicates that there is a significant fraction of the infrared power, probably originating in embedded star forming regions, that was not probed through the photospheric+nebular fit, in those sources.
\end{enumerate}

Deeper and additional data, especially in the near-infrared, are needed to constrain more accurately the mass of the various stellar populations as the youngest population tends the dominate the radiation field even when representing a small fraction of the total stellar mass. A tentative constraint of the IMF has been obtained on one object. More accurate data are required to get significant results   for the other regions in our sample.

Furthermore, the interstellar radiation field intensity cannot be determined accurately without data in the far-infrared. This makes a detailed study of the dust properties (such as the grain size distribution for instance) purely speculative. The recently launched Herschel Space Observatory is the key instrument to perform such a study, providing crucial far-infrared data while retaining a good spatial resolution to resolve star forming regions in collision debris from their nearby environment.

\acknowledgments

We would like to thank the anonymous referee for the useful comments that have helped improve and clarify this manuscript.

UL acknowledges financial support by the Spanish Science Ministry under grant AYA 2007-67625-C02-02 and by the Junta de Andaluc\'ia. VC would like to acknowledge partial support from the EU ToK grant 39965 and FP7-REGPOT 206469.

{\it Facilities:} \facility{Akari}, \facility{Spitzer}, \facility{GALEX}, \facility{CAO:2.2m}, \facility{CFHT}, \facility{ESO:3.6m}, \facility{KPNO:2.1m}, \facility{NTT}.
\bibliographystyle{aa}
\bibliography{article}

\end{document}